\documentclass[aps,prd,onecolumn,nofootinbib,groupedaddress,superscriptaddress,10pt]{revtex4}  
\usepackage{graphicx}
\usepackage{epstopdf}
\usepackage{amsmath}
\usepackage{amsfonts}
\usepackage{amssymb}
\usepackage{appendix}
\usepackage{comment}
\usepackage{bbold}
\usepackage{color}
\usepackage{slashed}
\usepackage{subfigure}
\usepackage{setspace}
\usepackage{footnote}
\usepackage{multirow}
\usepackage{longtable}
\usepackage{braket}
\usepackage[letterpaper, portrait, top=0.75in, left=0.5in, right=0.4in, bottom=0.4in]{geometry}

\newcommand{\beq}{\begin{equation}}
\newcommand{\eeq}{\end{equation}}

\usepackage{hyperref}
\definecolor{nicered}{rgb}{0.7,0.1,0.1}
\definecolor{nicegreen}{RGB}{53,170,102}
\definecolor{niceblue}{RGB}{57,94,221}
\definecolor{nicepurple}{RGB}{127, 38, 222}
\hypersetup{colorlinks,
			citecolor      = nicepurple, 
			linkcolor       = niceblue, 
			urlcolor        = nicegreen, 
			anchorcolor = niceblue}

\begin{document}

\singlespacing

{\hfill FERMILAB-PUB-22-247-T, CERN-TH-2022-060}

\title{Very Light Sterile Neutrinos at NOvA and T2K}

\author{Andr\'{e} de Gouv\^{e}a} 
\affiliation{Northwestern University, Department of Physics \& Astronomy, 2145 Sheridan Road, Evanston, IL 60208, USA}
\author{Giancarlo Jusino S\'anchez} 
\affiliation{Northwestern University, Department of Physics \& Astronomy, 2145 Sheridan Road, Evanston, IL 60208, USA}
\author{Kevin J. Kelly}
\affiliation{Theoretical Physics Department, Fermilab, P.O. Box 500, Batavia, IL 60510, USA}
\affiliation{Theoretical Physics Department, CERN, Esplande des Particules, 1211 Geneva 23, Switzerland}

\begin{abstract}
Over the last several years, our understanding of neutrino oscillations has developed significantly due to the long-baseline measurements of muon-neutrino disappearance and muon-to-electron-neutrino appearance at the T2K and NOvA experiments. However, when interpreted under the standard-three-massive-neutrinos paradigm, a tension has emerged between the two experiments' data. Here, we examine whether this tension can be alleviated when a fourth, very light neutrino is added to the picture. Specifically, we focus on the scenario in which this new neutrino has a mass similar to, or even lighter than, the three mostly-active neutrinos that have been identified to date. We find that, for some regions of parameter space, the four-neutrino framework is favored over the three-neutrino one with moderate ($\lesssim 2\sigma$) significance. Interpreting these results, we provide future outlook for near-term and long-term experiments if this four-neutrino framework is indeed true.
\end{abstract}


\maketitle

\section{Introduction}
\label{sec:Intro}
\setcounter{equation}{0}

Long-baseline neutrino oscillation experiments aim at studying the phenomenon of neutrino oscillations by taking advantage of the known neutrino oscillation lengths, proportional to (the inverse of) the mass-squared differences $\Delta m^2_{21}\equiv m_2^2-m_1^2$ or  $\Delta m^2_{31}\equiv m^3_2-m_1^2$, where $m_{1,2,3}$ are the masses of the neutrino mass eigenstates $\nu_{1,2,3}$, respectively. The neutrino masses are labelled such that $m_2^2>m_1^2$ and $|\Delta m^2_{31}|>\Delta m^2_{21}$. With this definition, the sign of $\Delta m^2_{31}$ is an observable and captures the neutrino-mass ordering: normal ordering (NO) when $\Delta m^2_{31}$ is positive, inverted ordering (IO) when $\Delta m^2_{31}$ is negative.

Among the objectives of long-baseline experiments is testing the standard-three-massive-neutrinos paradigm, which states that there are three neutrino mass eigenstates and that these interact via neutral-current and charged-current weak interactions. As far as the charged-current weak interactions are concerned, three orthogonal linear combinations of  $\nu_{1,2,3}$ couple to the $W$-boson and the charged leptons $\ell_{\alpha}$ ($\alpha=e,\mu,\tau$). In more detail, $\nu_{\alpha}=U_{\alpha i}\nu_i$ ($i=1,2,3$) couples to $\ell_{\alpha}$ and the $W$-boson, and $U_{\alpha i}$ are the elements of the unitary leptonic mixing matrix. On the other hand, assuming the standard-three-massive-neutrinos paradigm is correct,  long-baseline experiments are capable of measuring, sometimes with great precision, the neutrino oscillation parameters -- the parameters which define $U_{\alpha i}$ and the mass-squared differences. One way to test the standard-three-massive-neutrinos paradigm is to assume it is correct; measure the oscillation parameters using different oscillation processes or different experimental setups; and compare the results. If different measurements of the same quantity disagree at a high confidence level, we would claim the underlying formalism -- in this case the standard three-massive-neutrinos paradigm -- is deficient. 

Among the current generation of long-baseline experiments are the Tokai to Kamioka experiment (T2K)~\cite{T2K:2011ypd,T2K:2021xwb}, in Japan, and the NuMI Off-axis $\nu_e$ Appearance (NOvA) experiment~\cite{NOvA:2016kwd,NOvA:2021nfi}, in the United States. They are sensitive to several of the neutrino oscillation parameters, including some that are, at present, virtually unknown: the neutrino mass-ordering and the CP-odd parameter $\delta_{CP}$ that governs whether and how much CP-invariance is violated in the lepton sector. Data from T2K and NOvA have been analyzed assuming the standard-three-massive-neutrinos paradigm and have led to interesting measurements of the oscillation parameters. Just as interesting, perhaps, is the fact that there is some tension between T2K and NOvA data. 

The tension, which was first demonstrated by Refs.~\cite{T2KNu2020,NOvANu2020}, has been quantified and examined critically in the three-neutrino framework by various authors~\cite{Kelly:2020fkv,Esteban:2020cvm,deSalas:2020pgw,Capozzi:2021fjo}. In a little more detail, both T2K and NOvA measure electron-like and muon-like events associated to a pion decay-in-flight neutrino source ($\pi\to\mu\nu_{\mu}$). Measurements are performed at both near and far detectors and the detectors are exposed to both ``neutrino'' and ``antineutrino'' beams. With all this information, they can infer the $\nu_{\mu}$ and $\overline{\nu}_{\mu}$ survival probabilities $P(\nu_{\mu}\to\nu_{\mu})$ and $P(\overline{\nu}_{\mu}\to\overline{\nu}_{\mu})$, respectively, and the $\nu_e$ and $\overline{\nu}_e$ appearance probabilities $P(\nu_{\mu}\to\nu_{e})$ and $P(\overline{\nu}_{\mu}\to\overline{\nu}_{e})$, respectively. At T2K, typical neutrino energies are around 600~MeV and the baseline is 295~km. Typical NOvA energies are around 2~GeV and the baseline is 810~km. 

Assuming the standard-three-massive-neutrinos paradigm, the T2K and NOvA disappearance data are consistent but the appearance data, for both neutrinos and antineutrinos, are in disagreement. Within the NO, T2K prefers $\delta_{\rm CP}$ values close to $3\pi/2$.\footnote{We will use the convention that CP-violating phases are defined over $[0, 2\pi]$.} In contrast, when analyzed under the NO, NOvA data have no strong preference for any particular value of $\delta_{\rm CP}$, however, they disfavor the combination of $\delta_{\rm CP}$ and the mixing angle $\sin^2\theta_{23}$ preferred by T2K at roughly $2\sigma$ confidence. This tension may be addressed by instead considering the IO, where both experiments prefer $\delta_{\rm CP} \approx 3\pi/2$~\cite{Kelly:2020fkv,T2K:2021xwb,NOvA:2021nfi}. However, global fits to all neutrino oscillation data, including those from reactor antineutrino experiments~\cite{DayaBay:2018yms,RENO:2018dro,DoubleChooz:2019qbj}, prefer NO at ${\sim}2-3\sigma$~\cite{Esteban:2020cvm,deSalas:2020pgw,Capozzi:2021fjo,Jimenez:2022dkn}, leaving the T2K-NOvA tension unaddressed.

Whether the tension can be alleviated by the presence of physics beyond the standard-three-massive-neutrinos paradigm has also been the subject of intense exploration (see, for example, Refs.~\cite{Denton:2020uda,Miranda:2019ynh,Chatterjee:2020yak,Chatterjee:2020kkm,Forero:2021azc,Rahaman:2021leu,Rahaman:2022rfp}). Here, we would like to explore, in some detail, whether the tension between T2K and NOvA can be interpreted as evidence for new light neutrino states. This issue has been discussed before~\cite{Chatterjee:2020yak}, assuming the new neutrino state $\nu_4$ with mass $m_4$ is relatively heavy: $|\Delta m^2_{41}|\gg|\Delta m^2_{31}|$. Instead, here we concentrate on $|\Delta m^2_{41}|$ values that are ${\cal O}(|\Delta m^2_{31}|)$ or smaller, down to ${\cal O}(\Delta m^2_{21})$, and explore the full parameter space associated with the fourth neutrino. In Sec.~\ref{sec:FourFlavor}, we describe the four-neutrino oscillation formalism of interest. We also discuss how the existence of a light fourth neutrino may help alleviate the T2K--NOvA tension. In Sec.~\ref{sec:Simulation} we present our simulations of NOvA and T2K data and discuss how these are used, in Sec.~\ref{sec:Results}, to compare the standard-three-massive-neutrinos paradigm and the fourth-neutrino hypothesis. We present some concluding remarks in Sec.~\ref{sec:conclusion}. Some results are included in appendices: Appendix~\ref{app:DetailFit} includes detailed numerical results from our analyses, Appendix~\ref{app:LowDm41} presents an alternate, extremely-light sterile neutrino analysis, and Appendix~\ref{app:Pseudoexperiments} discusses some Monte Carlo studies of T2K, NOvA, and their combination in light of the sterile neutrino analyses.

\section{Four-Flavor Neutrino Oscillations}
\label{sec:FourFlavor}
\setcounter{equation}{0}

We assume there are four neutrino mass eigenstates $\nu_{1,2,3,4}$, and that these are related to the four interaction eigenstates $\nu_{e,\mu,\tau}$ and $\nu_s$ (where we assume the $\nu_s$ state does not participate in the weak interactions) via a $4\times 4$ unitary mixing matrix:
\begin{equation}
U=R(\theta_{34})R(\theta_{24}, \delta_{24})R(\theta_{14},\delta_{14})R(\theta_{23})R(\theta_{13},\delta_{13})R(\theta_{12}),
\label{eq:rotmatrices}
\end{equation}
where $R$ are $4\times 4$ rotation matrices in the $ij$-plane associated with a rotation angle $\theta_{ij}$. The nontrivial entries of the different $R$ in Eq.~\eqref{eq:rotmatrices} are given by
\[
R(\theta_{ij})=
\begin{pmatrix} 
c_{ij} & s_{ij}\\
-s_{ij} & c_{ij}\\
\end{pmatrix}
\hspace{20 pt}
R(\theta_{ij},\delta_{ij})=
\begin{pmatrix} 
c_{ij} & s_{ij} e^{-\delta_{ij}}\\
-s_{ij} e^{\delta_{ij}} & c_{ij}\\
\end{pmatrix},
\]
where $c_{ij}=\cos{\theta_{ij}}$ and $s_{ij}=\sin{\theta_{ij}}$. This extension to the standard-three-massive-neutrinos paradigm includes one more independent mass-squared difference and five new mixing parameters: three mixing angles $(\theta_{14}, \theta_{24}, \theta_{34})$ and two complex phases $(\delta_{14}, \delta_{24})$.

The $4\times 4$ mixing matrix is defined in such a way that, in the limit $\theta_{14}, \theta_{24}, \theta_{34}\to 0$, $\nu_4=\nu_s$ and $\nu_{1,2,3}$ are linear superpositions of only the active states $\nu_{e,\mu,\tau}$. In this limit, we recover the standard-three-massive-neutrinos paradigm. We will be interested in the case where $\theta_{14}, \theta_{24}, \theta_{34}$ are relatively small and will refer to $\nu_{1,2,3}$ as the mostly active states. The mostly active states will be defined in the usual way, including the ordering of their masses, which is either ``normal'' (NO) or ``inverted'' (IO), as discussed in Sec.~\ref{sec:Intro}. With this in mind, we define
\begin{equation}
  \Delta m^2_{4l}\equiv\begin{cases}
    m^2_4-m^2_1, & \text{if $m_1<m_3$ (NO)}\\
    m^2_4-m^2_3, & \text{if $m_3<m_1$ (IO)}
  \end{cases}.
  \label{eq:order}
\end{equation}

\begin{figure}[htbp]
\begin{center}
\includegraphics[width=0.6\linewidth]{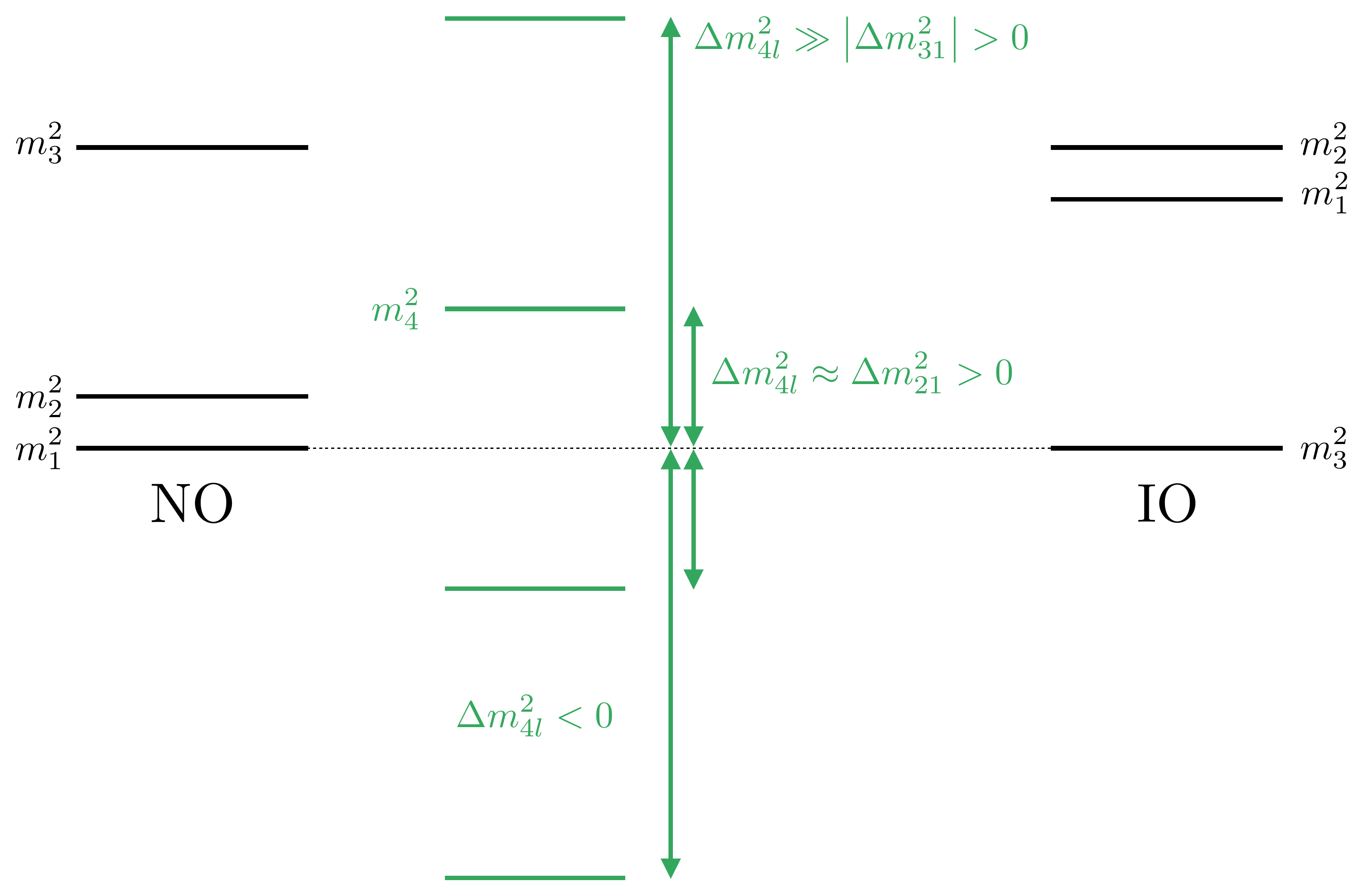}
\caption{Definition, including the sign convention, of $\Delta m^2_{4l}$ given the NO or IO for the mostly active states. \label{fig:MO}}
\end{center}
\end{figure}

In order to allow for all different relevant orderings of the four masses, we allow for both the NO and IO of the mostly active states and for both positive and negative values of $\Delta m^2_{4l}$. The four qualitatively different mass orderings are depicted in Fig.~\ref{fig:MO}. 
%
%
As far as the magnitude of $\Delta m^2_{4l}$, we will restrict our analyses to $(10^{-5} < |\Delta m^2_{4l}| < 10^{-1}) \text{ eV}^2$. Inside this range, we expect nontrivial oscillation effects to manifest themselves in the far detectors of T2K and NOvA but not in the corresponding near detectors. When $|\Delta m^2_{4l}|$ is smaller than $10^{-5}$~eV$^2$, the new oscillation length associated to $\Delta m^2_{4l}$ is too long and outside the reach of T2K and NOvA. Instead, when $|\Delta m^2_{4l}|$ is larger than $10^{-1}$~eV$^2$, we expect very fast oscillations in the far detectors of T2K and NOvA and nontrivial effects in the corresponding near detectors. This region of parameter space was explored in Ref.~\cite{Chatterjee:2020yak}.



The active neutrinos interact with the medium as they propagate from the source to the far detector. These interactions modify the equations that govern the flavor evolution of the neutrino states via effective potentials for forward charged-current (CC) and neutral-current (NC) scattering. The  neutrino flavor evolution equation can be written as a Schr\"odinger-like equation with an effective Hamiltonian given by, in the flavor basis, $H_F=1/(2E_{\nu})(U\textbf{M}^2U^{\dagger}+\textbf{A})$, where
\begin{equation}
\mathbf{M}^2=
\begin{pmatrix} 
0 & 0 & 0 & 0\\
0 & \Delta m_{21}^2 & 0 & 0\\
0 & 0 & \Delta m_{31}^2 & 0\\
0 & 0 & 0 & \Delta m_{41}^2\\
\end{pmatrix},
\hspace{20 pt}
\mathbf{A}=
\begin{pmatrix} 
2E_{\nu}V_{\rm CC} & 0 & 0 & 0\\
0 & 0 & 0 & 0\\
0 & 0 & 0 & 0\\
0 & 0 & 0 & -2E_{\nu}V_{\rm NC}\\
\end{pmatrix}.
\label{eq:mattermat}
\end{equation}
For neutrinos, $V_{\rm CC} = -2V_{\rm NC} = 3.8 \times10^{-5}\ (\text{eV}^2/\text{ GeV}) \rho [\frac{\text{g}}{\text{cm}^3}]$ are the CC and NC matter potentials, respectively. For antineutrinos, the matter potentials have the opposite sign. $\rho$ is the density -- assumed to be constant -- of the medium, assumed to be neutral. In this case, $V_{\rm NC}$ is half as large as $V_{\rm CC}$ and negative.
For the NOvA and T2K experiments, we fix the baselines to be $L_{\text{NOvA}}=810\text{ km}$ and $L_{\text{T2K}}=295\text{ km}$, respectively, while the near-far detector average matter densities are taken to be, respectively, $\rho_{\text{NOvA}}=2.8\ \text{g}/\text{cm}^3$~\cite{NOvA:2021nfi} and $\rho_{\text{T2K}}=2.6\ \text{g}/\text{cm}^3$~\cite{T2K:2021xwb}. The sterile nature of the new neutrino interaction eigenstate translates into a nontrivial $\mathbf{A}_{ss}$, obtained after the subtraction of $2E_{\nu}V_{\rm NC}\mathbb{1}$ from the Hamiltonian. 

\begin{figure}[htbp]
\begin{center}
\includegraphics[width=\linewidth]{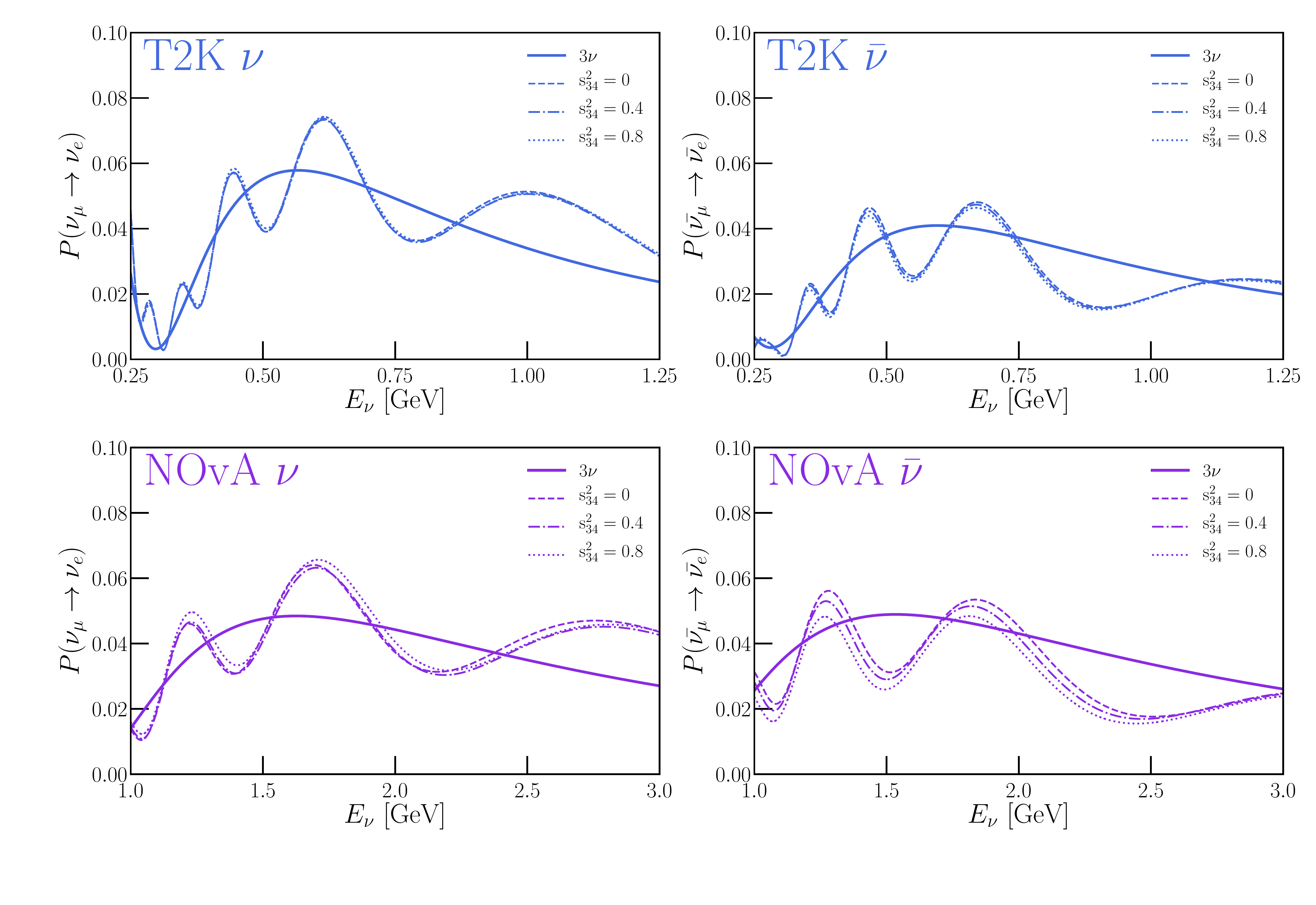}
\caption{Appearance oscillation probabilities at T2K (top, blue) and NOvA (bottom, purple) comparing three-neutrino oscillation probabilities (solid lines, parameters from Table \ref{tab:OscParams}, column 2 ``$3\nu$ IO'') against four-neutrino ones (non-solid lines, parameters from Table \ref{tab:OscParams}, column 4 ``4$\nu$ IO''). Left panels show probabilities for neutrino oscillation, whereas right ones show antineutrino oscillation. For the four-neutrino probabilities, three choices of $\sin^2{\theta_{34}}$ are used for illustrative purposes: dashed/dot-dashed/dotted lines correspond to $\sin^2{\theta_{34}}=0/0.4/0.8$.
\label{fig:plainprobs}}
\end{center}
\end{figure}

Since the tension between T2K and NOvA is mostly driven by the $\nu_e$ appearance channel, Fig.~\ref{fig:plainprobs} depicts the $\nu_e$ appearance probability for both experiments given the three-neutrino and four-neutrino hypotheses. The mixing parameters for the different hypotheses are listed in Table~\ref{tab:OscParams}, except for $\sin^2\theta_{34}$. We see that the new oscillation frequency $\left|\Delta m^2_{4l}\right| \approx 10^{-2}$ eV$^2$ can lead to pronounced oscillations at both NOvA and T2K. We also note that the new effects can be different at T2K relative to NOvA for, roughly, two different reasons. One is that the dominant values of $L/E$, keeping in mind that both beams have a narrow energy profile, are not identical for the two experiments. This means that for relatively ``fast'' $\Delta m^2_{4l}$ the value of the new oscillation phase will not be the same for the two experiments. The other is that the matter effects are more pronounced at NOvA relative to T2K. These allow the effective oscillation frequencies and mixing parameters to be distinct at the two experimental setups.

In vacuum, $P(\nu_{\mu}\to\nu_e)$ does not depend on $\theta_{34}$; this is not the case in matter. An easy way to see this is to express the propagation Hamiltonian in the mass basis. In the absence of matter effects, the dependency on the mixing parameters is encoded in the initial and final interaction eigenstates and since neither $\nu_e$ nor $\nu_{\mu}$, when expressed as linear superpositions of the mass eigenstates, depend on $\theta_{34}$, then neither can $P(\nu_{\mu}\to\nu_e)$. Instead, when the matter effects are present, the matter potential in the mass basis depends on $\theta_{34}$. Hence we expect $P(\nu_{\mu}\to\nu_e)$ to also depend on $\theta_{34}$ as long as matter effects are relevant. The dependency on $\theta_{34}$ can be seen in Fig. \ref{fig:plainprobs}. As expected, it is rather small at T2K and larger at NOvA, where matter effects are relatively more pronounced. 

\begin{figure}[htbp]
\begin{center}
\includegraphics[width=\linewidth]{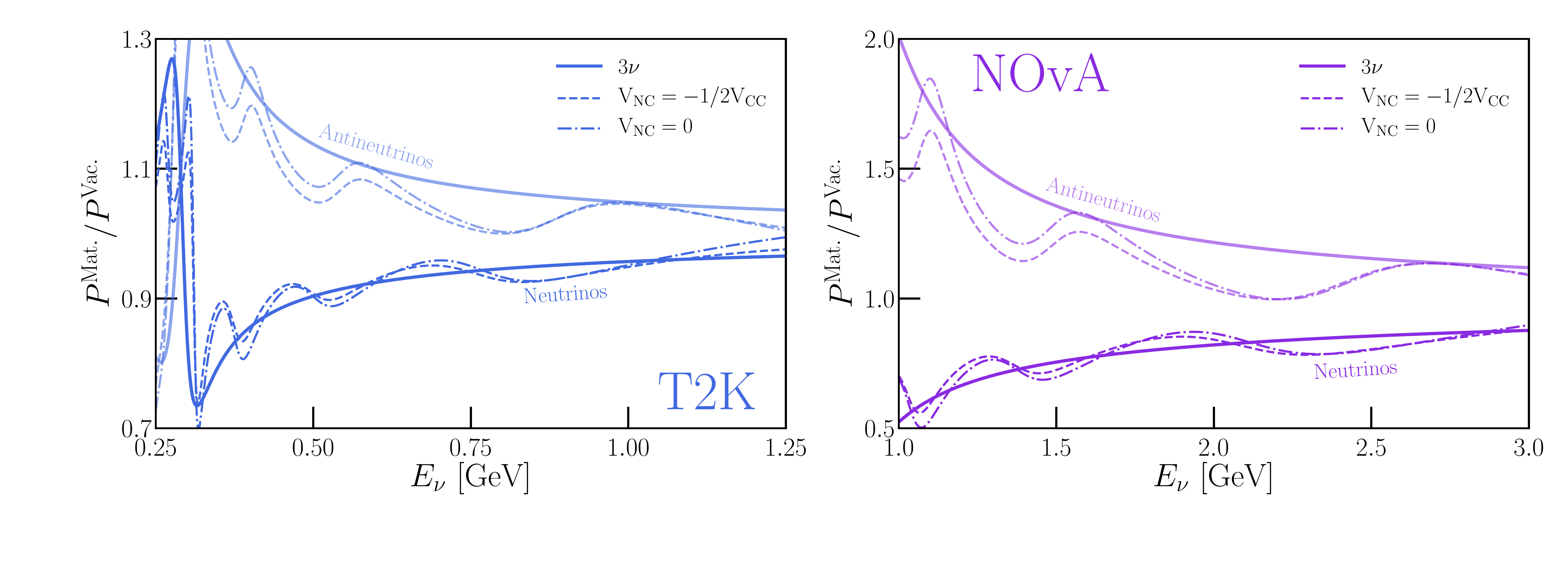}
\caption{Ratio of appearance oscillation probabilities in matter to those in vacuum at T2K (left) and NOvA (right). Solid lines correspond to the three-neutrino oscillation probabilities. Dashed and dot-dashed lines correspond to a fourth neutrino that is sterile or active, respectively. Parameters are taken from columns 2 and 4 from Table \ref{tab:OscParams} corresponding to the three-neutrino and four-neutrino cases, respectively. 
\label{fig:ratioprobs}}
\end{center}
\end{figure}

In order to further illustrate the impact of matter effects, Fig.~\ref{fig:ratioprobs} depicts the ratio of the appearance probabilities in matter relative to what those would be in vacuum. We also illustrate the difference between a new interaction state that is active and one that is sterile by depicting the same ratio assuming the neutral current matter potential is zero. The new oscillation frequency is apparent at both experiments and it is easy to see that matter effects are more pronounced at NOvA relative to T2K. The ``sterileness'' of the fourth neutrino is also more pronounced at NOvA, as expected.

\section{Simulating Data from NOvA and T2K}
\label{sec:Simulation}
\setcounter{equation}{0}
As discussed earlier, both NOvA and T2K operate with beams with a flux of predominantly $\nu_\mu$ ($\overline\nu_\mu$) when operating in (anti)neutrino mode. Both experiments' far detectors are designed to study the disappearance of $\nu_\mu$ and $\overline\nu_\mu$, as well as the appearance of $\nu_e$ and $\overline\nu_e$. Using the most recent publications from NOvA~\cite{NOvA:2021nfi} and T2K~\cite{T2K:2021xwb}, and building off the simulations of Refs.~\cite{Ellis:2020ehi,Ellis:2020hus,Kelly:2020fkv}, we perform simulations to determine the expected event rates in the disappearance and appearance channels of both experiments given a set of three- or four-neutrino oscillation parameters. We then compare these expected event rates against the experiments' published event rates and construct a test statistic using Poissonian bin expectations.

In the remainder of this section, we briefly explain the process by which we simulate the expected event rates, as well as the number of data points for each experiment that enter our test statistic. 
\begin{table}
\begin{center}
\caption{Oscillation parameters assumed when depicting oscillation probabilities and expected event rates. The four columns correspond to the three-neutrino ($3\nu$) and four-neutrino ($4\nu$) hypotheses, as well as whether the three mostly-active neutrinos follow the normal (NO) or inverted (IO) mass ordering.\label{tab:OscParams}}
\begin{tabular}{|c||c|c||c|c|}\hline
Parameter & $3\nu$ NO & $3\nu$ IO & $4\nu$ NO & $4\nu$ IO \\ \hline\hline
$\sin^2 \theta_{12}$ & 0.307 & 0.307 & 0.321 & 0.314 \\ \hline
$\sin^2 \theta_{13}$ & 0.022 & 0.022 & $0.023$ & $0.023$ \\ \hline
$\sin^2\theta_{23}$ & 0.57 & 0.57 & $0.43$ & $0.45$ \\ \hline
$\Delta m_{21}^2/10^{-5}$ eV$^2$ & 7.53 & 7.53 & 7.53 & 7.53 \\ \hline
$\Delta m_{31}^2/10^{-3}$ eV$^2$ & 2.51 & -2.41 & 2.49 & -2.39 \\ \hline
$\delta_{\rm CP}$ & 3.66 & 4.71 & 4.09 & 4.46 \\ \hline\hline
$\sin^2\theta_{14}$ & --- & --- & 0.043 & 0.021 \\ \hline
$\sin^2\theta_{24}$ & --- & --- & 0.060 & 0.053 \\ \hline
$\sin^2\theta_{34}$ & --- & --- & 0.37 & 0.56 \\ \hline
$\Delta m_{41}^2$/eV$^2$ & --- & --- & $1.1 \times 10^{-2}$ & $-1.1\times 10^{-2}$ \\ \hline
$\delta_{14}$ & --- & --- & 0.01 & 4.88 \\ \hline
$\delta_{24}$ & --- & --- & 1.82 & 5.89 \\ \hline\hline
\end{tabular}
\end{center}
\end{table}
To center our discussion, we will rely on several benchmark sets of oscillation parameters with which we calculate the expected observables at NOvA and T2K. We adopt two benchmark sets each for the $3\nu$ and $4\nu$ assumptions, listed in Table~\ref{tab:OscParams}, allowing for the mostly-active neutrinos to follow either the normal (NO) or inverted (IO) orderings. As we will discuss in Section~\ref{sec:Results}, these parameters are the best-fit points obtained by our fit to the \textit{combination} of T2K and NOvA under the different hypotheses.

\textbf{NOvA ---} Our simulation of NOvA, designed to match the results of Ref.~\cite{NOvA:2021nfi}, includes the disappearance channels of neutrino and antineutrino mode (19 bins each, with neutrino energies ranging from $0$ to $5$ GeV) as well as event rate measurements of the appearance channels\footnote{For simplicity, we sum the expected event rate for the entire neutrino energy range and compare it against the observed 82 (33) appearance events of operation in (anti)neutrino mode.}, totaling 40 data points. This simulation corresponds to a total exposure of $13.6 \times 10^{20}$ ($12.5 \times 10^{20}$) protons on target (POT) in (anti)neutrino mode.

\begin{figure}[htbp]
\begin{center}
\includegraphics[width=\linewidth]{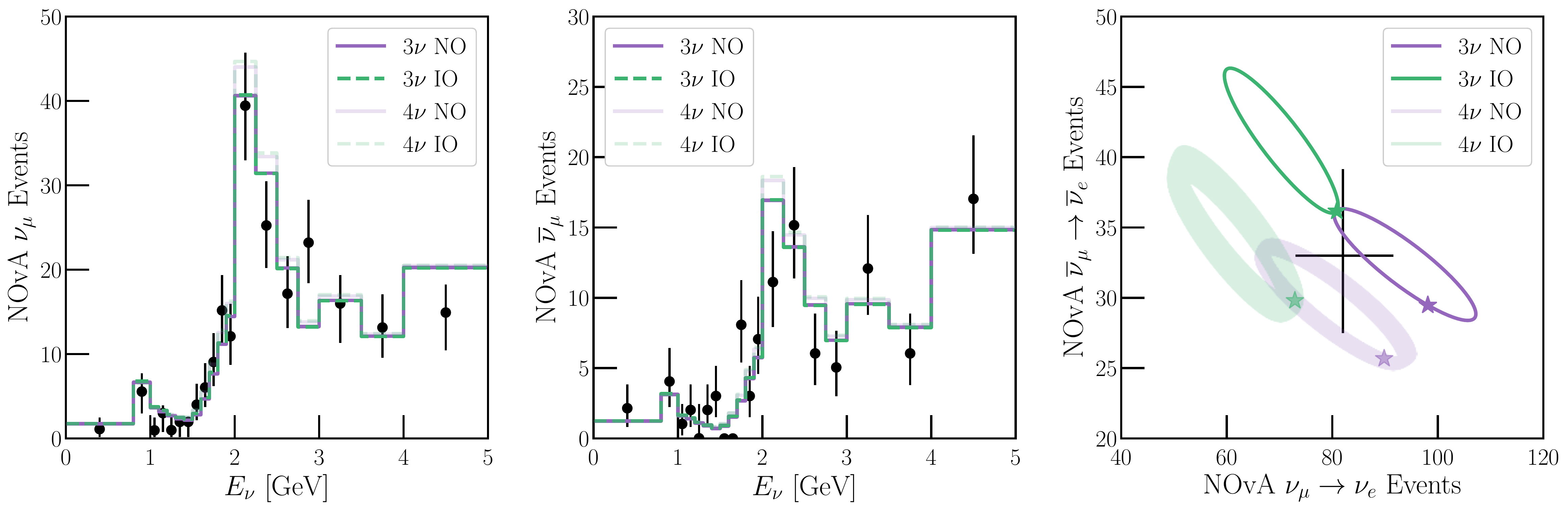}
\caption{Expected and observed event rates in NOvA's $\nu_\mu$ disappearance (left), $\overline\nu_\mu$ disappearance (center), and $\nu_e$/$\overline\nu_e$ appearance (right) channels. We compare the prediction under the $3\nu$ (solid/dashed lines) and $4\nu$ (faint lines/regions) hypotheses, with parameters from Table~\ref{tab:OscParams}, with the observed data (black). Purple curves correspond to the mostly-active neutrinos following the normal mass ordering (NO), where green ones correspond to the inverted mass ordering (IO). In the right panel, the CP-violating phases are allowed to vary in the predicted rates. Data points from Ref.~\cite{NOvA:2021nfi}.\label{fig:NOVAValidation}}
\end{center}
\end{figure}
Fig.~\ref{fig:NOVAValidation} shows the expected event rates in NOvA for neutrino mode $\nu_\mu$ disappearance (left), antineutrino mode $\overline\nu_\mu$ disappearance\footnote{In contrast to Ref.~\cite{NOvA:2021nfi}, our disappearance channel panels depict the event rate per bin as opposed to event rate per unit energy, causing our higher-energy bins (with larger bin width) to appear exaggerated.} (center), and a joint comparison of neutrino ($x$-axis) and antineutrino ($y$-axis) mode $\nu_\mu \to \nu_e$ (or $\overline\nu_\mu \to \overline\nu_e$) appearance (right panel). We compare the NOvA benchmark oscillation predictions, using the parameters in Table~\ref{tab:OscParams} (purple histograms/curves\footnote{Where the faint curves are not visible in the left/center panels, the four-neutrino hypothesis predicts the same rate as the three-neutrino one(s).} for NO, green for IO, and dark curves for $3\nu$, faint ones for $4\nu$), to the observed event rates from the experiment (black). Error bars here are only statistical. In the left and center panels, all oscillation parameters are fixed according to Table~\ref{tab:OscParams}. In contrast, the right panel allows $\delta_{\rm CP}$ to vary for the $3\nu$ curves, and all three CP-violating phases to vary in the $4\nu$ case. This allows for a set of ellipses in this bi-event parameter space instead of a single one. In the right panel, stars indicate the predicted event rates when the CP-violating phases are fixed to their values in Table~\ref{tab:OscParams}.

\textbf{T2K ---} We simulate T2K in much the same spirit as NOvA, with the goal of matching the results presented in Ref.~\cite{T2K:2021xwb}. In the case of T2K, the disappearance channels each consist of 30 bins -- $100$ MeV in width from $0$ to $2.9$ GeV, and one bin corresponding to neutrino energies above $2.9$ GeV. For the appearance channel, we take advantage of the expected neutrino-energy spectrum with bins of $125$ MeV width from $0$ to $1.25$ GeV in each channel.\footnote{Refs.~\cite{Ellis:2020ehi,Ellis:2020hus}, however, have demonstrated that total-rate measurements of T2K's appearance channel result in similar parameter estimation to the collaboration's results.} This yields 80 data points in our T2K analysis. Our T2K simulation corresponds to an exposure of $14.94 \times 10^{20}$ ($16.35 \times 10^{20}$) POT in (anti)neutrino mode operation.
\begin{figure}
\begin{center}
\includegraphics[width=\linewidth]{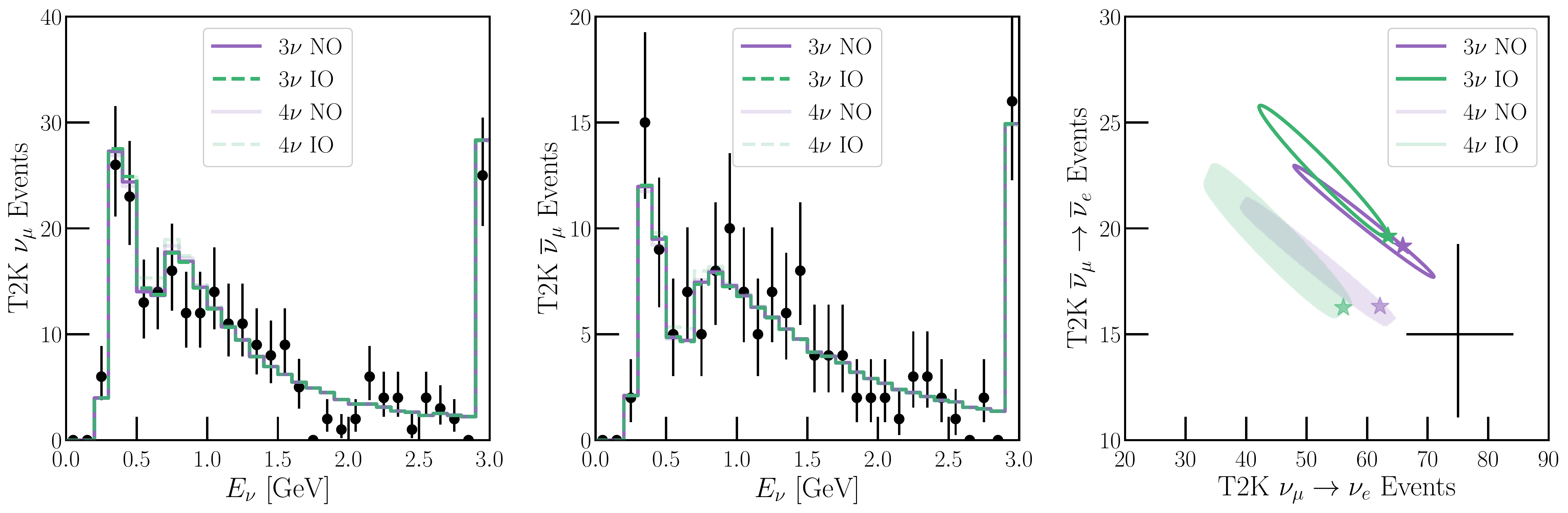}
\caption{Expected and observed event rates in T2K's $\nu_\mu$ disappearance (left), $\overline\nu_\mu$ disappearance (center), and $\nu_e$/$\overline\nu_e$ appearance (right) channels. We compare the prediction under the $3\nu$ (solid/dashed lines) and $4\nu$ (faint lines/regions) hypotheses, with parameters from Table~\ref{tab:OscParams}, with the observed data (black). Purple curves correspond to the mostly-active neutrinos following the normal mass ordering (NO), where green ones correspond to the inverted mass ordering (IO). In the right panel, the CP-violating phases are allowed to vary in the predicted rates. Data points from Ref.~\cite{T2K:2021xwb}.\label{fig:T2KValidation}}
\end{center}
\end{figure}

Similar to Fig.~\ref{fig:NOVAValidation}, we show in Fig.~\ref{fig:T2KValidation} our expected event rates in the different T2K channels -- the left panel is for $\nu_\mu$ disappearance, center for $\overline\nu_\mu$ disappearance, and the right panel is the combined $\nu_e$ and $\overline\nu_e$ appearance. For clarity of display, we sum the total expected event rates in the $\nu_e$ and $\overline\nu_e$ channels in the right panel. Here, the oscillation parameters correspond to those given in Table~\ref{tab:OscParams} and, in the right panel, the CP-violating phases are allowed to vary.

\textbf{Test Statistic ---} We take the expected and observed event rates in NOvA (40 data points), T2K (80), or a combination of them (120) and construct a test statistic using Poisson statistics for the log-likelihood (matching a $\chi^2$ function in the limit of large event rates):
\begin{equation}\label{eq:chi2}
\chi^2 = \sum_{i\ \in\ \mathrm{bins}} -2\left(-\lambda_i + x_i + x_i\log{\left(\frac{\lambda_i}{x_i}\right)}\right),
\end{equation}
where $\lambda_i$ ($x_i$) represents the expected (observed) event rate in bin $i$ for a given experiment/channel.

We will be interested in several pieces of information from the test statistic in Eq.~\eqref{eq:chi2}. When performing parameter estimations, we will use contours of $\Delta\chi^2$ about its minimum to represent preferred regions/intervals of parameter space. When comparing best-fit points under different hypotheses, i.e., comparing preference for the $4\nu$ scenario over the $3\nu$ one, we will compare the minimum $\chi^2$ when varying over oscillation parameters, taking into account the number of degrees of freedom in such a fit. 

\textbf{Analysis \& Priors ---} The main focus of this work is on the long-baseline experiments NOvA and T2K, which are sensitive to oscillation effects associated with mass-squared differences of order of $10^{-3}$ eV$^2$. On the other hand, the solar mass-squared difference has been well-measured by solar neutrino~\cite{Super-Kamiokande:2016yck,SKNu2020} and reactor antineutrino~\cite{KamLAND:2013rgu} experiments to be $\Delta m_{21}^2 = 7.53 \times 10^{-5}$ eV$^2$ while the associated mixing angle is measured to be $\sin^2\theta_{12} = 0.307$, both at the few percent level. Due to the lack of sensitivity to these quantities at NOvA/T2K, we fix them\footnote{Specifically, we fix the matrix-element-squared $|U_{e2}|^2$, which is equal to $\sin^2\theta_{12} \cos^2\theta_{13} \cos^2\theta_{14}$ in the four-neutrino framework, to its best-fit value of $0.300$. This causes $\sin^2\theta_{12}$ to vary for large $\theta_{14}$.} in our analyses. While NOvA and T2K are sensitive to $\sin^2\theta_{13}$ through their appearance channels, their measurement capability is significantly weaker than that of Daya Bay~\cite{DayaBay:2018yms}, RENO~\cite{RENO:2018dro}, and Double Chooz~\cite{DoubleChooz:2019qbj} reactor antineutrino experiments. In our fits, we include Daya Bay's measurement as a Gaussian prior on the quantity $4|U_{e3}|^2 ( 1 - |U_{e3}|^2) = 0.0856 \pm 0.0029$, which is $\sin^2(2\theta_{13})$ when considering the three-neutrino hypothesis~\cite{DayaBay:2018yms}.

\section{Results}
\label{sec:Results}
\setcounter{equation}{0}
\setcounter{footnote}{0}
This section details the results of our analyses. First, in Section~\ref{subsec:3nuResults}, we summarize the results of fits of our NOvA and T2K simulations and their combination under the three-neutrino hypothesis. Then, Section~\ref{subsec:4nuResults} discusses the results of these fits under the four-neutrino hypothesis, including a comparison of the three-neutrino and four-neutrino hypotheses.

\subsection{Three-Neutrino Results}\label{subsec:3nuResults}
Our first three-neutrino analysis is focused on finding the best-fit points of each experimental analysis (T2K, NOvA, and a combined fit). For this, we perform two fits for each experiment/combination, one assuming that neutrinos follow the normal mass ordering (NO, $\Delta m_{31}^2 > 0$) and one assuming that they follow the inverted one (IO, $\Delta m_{31}^2 < 0$). Recent results have demonstrated that, under the three-neutrino hypothesis, T2K and NOvA each exhibit mild preference for the NO over the IO, but their combination has a mild preference for the IO~\cite{Kelly:2020fkv,Esteban:2020cvm,deSalas:2020pgw,Capozzi:2021fjo}. When combined with all reactor antineutrino data and other experimental results, the global preference is for the NO at relatively low significance.

We find a result consistent with these previous results, summarized in Table~\ref{tab:BFPs3nu}. As in all of our analyses, $\Delta m_{21}^2$ and $\sin^2\theta_{12}$ are fixed, and a prior is included from the results of Daya Bay on $\sin^2(2\theta_{13})$. We present both the overall test statistic at this best-fit point for each analysis as well as the preference for the NO over the IO in the right-most column (positive values indicate preference for NO, negative for IO). We note here that all of the best-fit $\chi^2$ obtained are comparable to (and in the case of T2K and the joint fit, less than) the number of degrees of freedom, implying that these are all good fits to their respective data sets. Finally, we see that the joint-fit $\chi^2$ under the NO hypothesis is around five units of $\chi^2$ larger than the sum of the two individual fits whereas, under the IO hypothesis, it is roughly the same -- this highlights the so-called NOvA/T2K tension, where the results disagree under the NO hypothesis but not under the IO one. The values from the ``Joint'' fit in Table~\ref{tab:BFPs3nu} correspond to the benchmark values we adopted in the three-neutrino case in Table~\ref{tab:OscParams}.
\begin{table}[!htbp]
\begin{center}
\caption{Best-fit parameters of our analyses of T2K, NOvA, and a combined analysis of the two under the three-neutrino hypothesis. We determine the best-fit point under the normal (NO) and inverted (IO) mass-ordering hypotheses, as well as the overall preference for the NO over IO, $\Delta \chi^2_{\rm NO,IO}$, for each analysis. In each, a prior on $\sin^2(2\theta_{13})$ from Daya Bay is included, and $\sin^2\theta_{12} = 0.307$ and $\Delta m_{21}^2 = 7.53 \times 10^{-5}$ eV$^2$ are fixed to their best-fit points from other experimental results.\label{tab:BFPs3nu}}
\begin{tabular}{c| c||c|c|c|c||c||c}
\multicolumn{2}{c||}{3$\nu$} & $\sin^2\theta_{13}$ & $\sin^2\theta_{23}$ & $\Delta m_{31}^2/10^{-3}$ eV$^2$ & $\delta_{\rm CP}$ & $\chi^2$ & $\Delta \chi^2_{\rm NO,IO}$ \\ \hline \hline
\multirow{2}{*}{T2K} & NO & $0.022$ & $0.56$ & $2.52$ & $4.58$ & $66.82$ & \multirow{2}{*}{$1.48$} \\ \cline{2-7}
& IO & $0.022$ & $0.56$ & $-2.41$ & $4.71$ & $68.19$ & \\ \hline
\multirow{2}{*}{NOvA} & NO & $0.022$ & $0.58$ & $2.52$ & $2.34$ & $43.40$ & \multirow{2}{*}{$0.14$} \\ \cline{2-7}
& IO & $0.022$ & $0.57$ & $-2.41$ & $4.78$ & $43.55$ & \\ \hline
\multirow{2}{*}{Joint} & NO & $0.022$ & $0.57$ & $2.51$ & $3.67$ & $115.58$ & \multirow{2}{*}{$-3.76$} \\ \cline{2-7}
& IO & $0.022$ & $0.57$ & $-2.41$ & $4.72$ & $111.82$ & \\ \hline
\end{tabular}
\end{center}
\end{table}

We also perform a parameter estimation under the three-neutrino hypothesis, both to prepare our expectations for the four-neutrino analyses and to validate our results compared against the official results of the experimental collaborations. The free/fixed parameters and test statistic are identical to those when determining the best-fit points. For simplicity, we perform an analysis of the parameters $\sin^2\theta_{13}$, $\sin^2\theta_{23}$, $\Delta m_{31}^2$, and $\delta_{\rm CP}$ and marginalize over $\sin^2\theta_{13}$ and $\Delta m_{31}^2$ (including both the NO and IO hypotheses), and present the joint measurement of $\sin^2\theta_{23}$ and $\delta_{\rm CP}$.
\begin{figure}[!htbp]
\begin{center}
\includegraphics[width=0.6\linewidth]{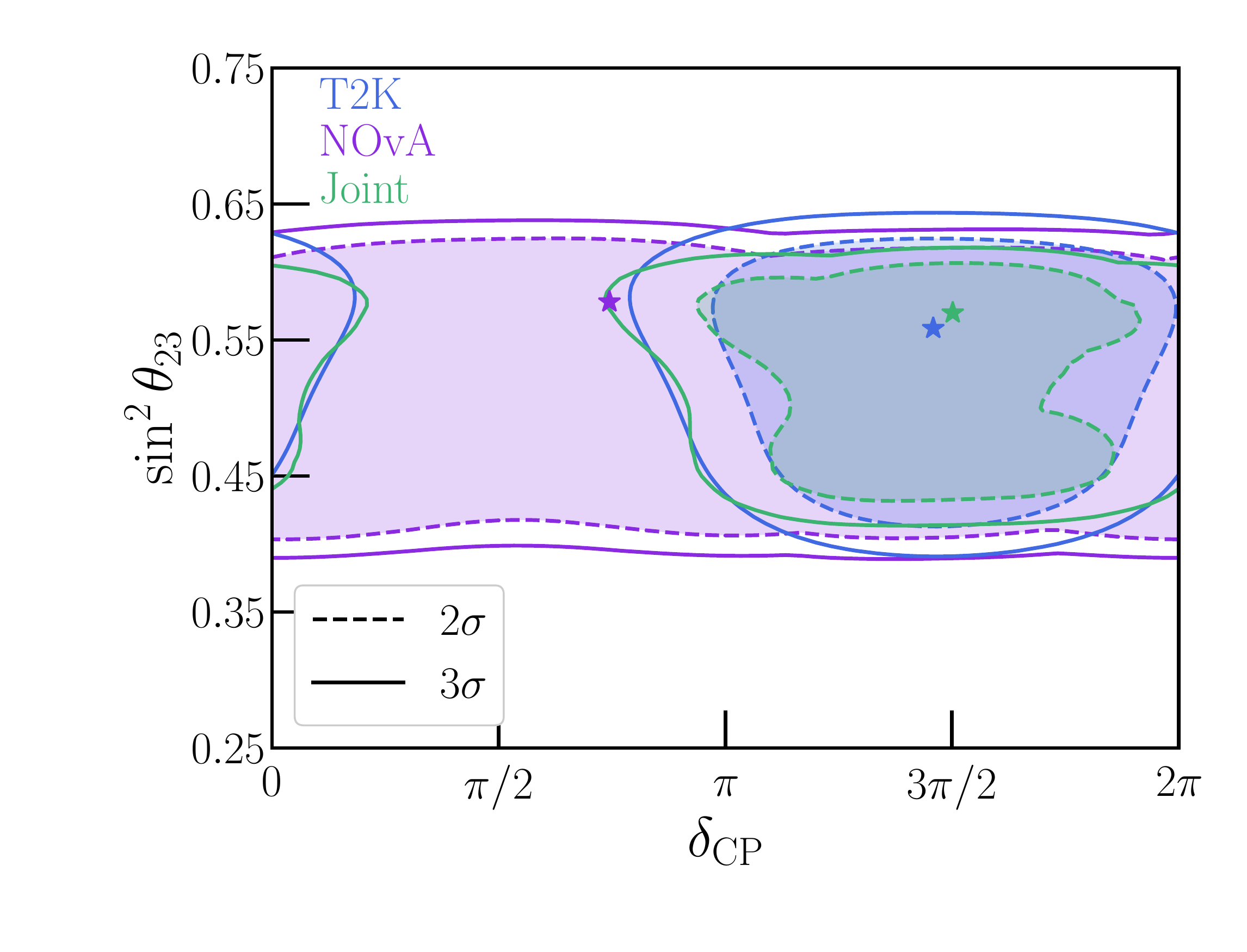}
\caption{Parameter estimation of $\delta_{\rm CP}$ and $\sin^2\theta_{23}$ from T2K (blue), NOvA (purple), and their combination (green) at $2\sigma$ (dashed lines) and $3\sigma$ (solid lines) CL. \label{fig:ThreeNeutrinoScan}}
\end{center}
\end{figure}

Fig.~\ref{fig:ThreeNeutrinoScan} presents the results of this analysis at $2\sigma$ (dashed, filled contours) and $3\sigma$ (solid lines) CL for T2K (blue), NOvA (purple), and the joint fit (green). Stars of each color represent the best-fit points obtained in Table~\ref{tab:BFPs3nu}. Once the mass ordering is marginalized, NOvA has no sensitivity to $\delta_{\rm CP}$, and constrains $\sin^2\theta_{23}$ to be between roughly $0.37$ and $0.65$ at $3\sigma$ CL. In the NO, NOvA can take on nearly any value of $\delta_{CP}$, however it disfavors the combination $\delta_{CP} = 3\pi/2$, $\sin^2\theta_{23} > 1/2$ at relatively high significance. Under the IO, NOvA prefers this combination. Regardless of the mass ordering, T2K prefers $\delta_{\rm CP} = 3\pi/2$ and constrains $\sin^2\theta_{23}$ to be in a similar range as NOvA. When the two are combined, the preferred regions are very similar to those obtained in the fit to T2K data alone.

\subsection{Four-Neutrino Results}\label{subsec:4nuResults}

We begin our four-neutrino analyses by repeating the process that led to Table~\ref{tab:BFPs3nu} -- we determine the best-fit points under the four-neutrino hypothesis for T2K, NOvA, and their combination. Now that we are considering four-neutrino oscillations, we allow for all four mass orderings discussed in Sec.~\ref{sec:FourFlavor} (see Fig.~\ref{fig:MO}). This amounts to dividing the analysis based on the signs of $\Delta m_{31}^2$ and $\Delta m_{4l}^2$, where $l$ represents $m_1$ in the NO and $m_3$ in the IO, the lightest of the mostly-active neutrinos.

Table~\ref{tab:BFPs4nuSmall} summarizes these twelve analyses (four each for NOvA, T2K, and their Joint fit), giving the best-fit parameters as well as the overall $\chi^2$ of each fit in the four-neutrino hypothesis.  Near the bottom we give the preferred ordering of masses from each experiment/combination -- T2K and the Joint fit both prefer $m_4 < m_3 < m_1 < m_2$, where NOvA prefers $m_1 < m_2 < m _3 < m_4$.  The preference for the sign of $\Delta m_{4l}^2$ is small in all cases -- individual fit results for all four mass orderings and all three experimental combinations are provided for completeness in Appendix~\ref{app:DetailFit}. When allowing for a fourth neutrino, neither T2K nor NOvA have a strong preference for the sign of $\Delta m_{31}^2$. T2K prefers $\Delta m_{31}^2 < 0$ at $\Delta \chi^2 = 0.1$, where NOvA prefers $\Delta m_{31}^2 > 0$ at $\Delta \chi^2 = 0.02$. However, the combined fit prefers $\Delta m_{31}^2 < 0$ at $\Delta \chi^2 = 4.6$ an even stronger preference for negative $\Delta m^2_{31}$ than when data are analyzed under the three-neutrino hypothesis.
\begin{table}
\begin{center}
\caption{Best-fit parameters of the four-neutrino analyses of T2K, NOvA, and their combination. We allow for all possible orderings of the neutrino mass eigenstates, hence $\Delta m_{31}^2$ and $\Delta m_{4l}^2$ can each be negative. In each analysis, a prior on $|U_{e3}|^2(1-|U_{e3}|^2)$ from Daya Bay is included, and $|U_{e2}|^2 = 0.300$ and $\Delta m_{21}^2 = 7.53 \times 10^{-5}$ eV$^2$ are fixed to their best-fit points from other experimental results.\label{tab:BFPs4nuSmall}}
\begin{tabular}{c||c|c|c|}
4$\nu$ & T2K & NOvA & Joint \\ \hline \hline
$\sin^2\theta_{13}$ & $0.024$ & $0.022$ & $0.023$ \\ \hline
$\sin^2\theta_{23}$ & $0.43$ & $0.44$ & $0.43$ \\ \hline
$\Delta m_{31}^2/10^{-3}$ eV$^2$ & $-2.39$ & $2.43$ & $-2.39$ \\ \hline
$\delta_{\rm CP}$ & $4.41$ & $0.00$ & $4.46$ \\ \hline\hline
$\sin^2\theta_{14}$ & $7.8\times 10^{-2}$ &  $6.9\times 10^{-3}$ &  $4.3\times 10^{-2}$ \\ \hline
$\sin^2\theta_{24}$ & $4.1\times 10^{-2}$ &  $1.2\times 10^{-1}$ &  $6.0\times 10^{-2}$ \\ \hline
$\sin^2\theta_{34}$ & $0.78$ &  $0.29$ &  $0.37$ \\ \hline
$\Delta m_{4l}^2/$eV$^2$ & $-8.5\times10^{-3}$ & $1.0\times10^{-2}$ & $-8.5\times10^{-3}$ \\ \hline
$\delta_{14}$ & $1.82$ & $3.51$ & $4.88$ \\ \hline
$\delta_{24}$ & $2.64$ & $3.15$ & $5.89$ \\ \hline \hline
$\chi^2_{\rm 4\nu}$ & $61.95$ & $38.10$ & $102.83$ \\ \hline
Ordering & $m_4 < m_3 < m_1 < m_2$ & $m_1 < m_2 < m_3 < m_4$ & $m_4 < m_3 < m_1 < m_2$ \\ \hline
$\chi^2_{\rm 3\nu} - \chi^2_{\rm 4\nu}$ & $4.87$ & $5.30$ & $8.99$
\end{tabular}
\end{center}
\end{table}

The bottom row of Table~\ref{tab:BFPs4nuSmall} presents the improvement in each experimental analysis (as well as the combined one) compared to the results of the three-neutrino analysis. We find that the fits to both the T2K\footnote{This result is consistent with what the T2K collaboration reported in Ref.~\cite{T2K:2019efw}, which found an improvement of $\Delta \chi^2 = 4.7$.} and NOvA data improve by roughly five units in $\chi^2$, and the combined fit improves by nearly nine units. However, we note two very important caveats here:
\begin{enumerate}
\item The results of the three-neutrino fit in Table~\ref{tab:BFPs3nu} demonstrate that, relative to the number of degrees of freedom, good fits have been achieved. So, when comparing the three-neutrino fit -- four parameters --  to the four-neutrino one -- ten parameters --  one must take into account the fact that this minimization is being performed over an additional six parameters.
\item When determining the statistical significance, the comparison of $\chi^2_{3\nu} - \chi^2_{4\nu}$ must be scrutinized to see whether these test statistics follow a $\chi^2$ distribution. We have performed some basic Monte Carlo studies of our T2K and NOvA simulations (see Appendix~\ref{app:Pseudoexperiments}) and found that, when statistical fluctuations are considered, one will often find best-fit points with $\Delta m_{4l}^2 \approx 10^{-2}$ eV$^2$ that improve each experiment's fit by a couple of units of $\chi^2$. This is likely driven by the sizes of the energy bins (around 100~MeV) used in the T2K and NOvA analyses -- at T2K/NOvA baselines/energies, a new oscillation driven by a mass-squared splitting of $10^{-2}$ eV$^2$ will evolve significantly\footnote{For this $\Delta m^2$, the argument of the term $\sin^2(\Delta m^2 L/4E_\nu)$ that enters the oscillation probabilities changes by an appreciable fraction of $\pi$.} over the span of a single bin. This new fast oscillation can ``absorb'' individual bins' statistical fluctuations and lead to an artificial improvement in the test statistic. This is validated by the results of Ref.~\cite{T2K:2019efw}, which found that an improvement of $\Delta \chi^2 = 4.7$ at T2K (between the three-neutrino and four-neutrino hypotheses) corresponds to only ${\sim}1.0\sigma$ preference for a fourth neutrino, in contrast with the preference derived assuming Wilks' theorem~\cite{Wilks:1938dza} holds, ${\sim}1.7\sigma$.
\end{enumerate}
When considering the results of Table~\ref{tab:BFPs4nuSmall} (and that the best-fit points are close to $|\Delta m_{4l}^2| \approx 10^{-2}$ eV$^2$) in light of these two caveats, we find that, while a very light sterile neutrino improves the ``tension'' between T2K and NOvA, there is not strong evidence in favor of a four-neutrino hypothesis over the three-neutrino one.

In order to determine whether the sterile neutrino solution to the NOvA/T2K tension persists in light of caveat 2 above, we also perform an alternate analysis in Appendix~\ref{app:LowDm41} where we restrict $\Delta m_{21}^2 \lesssim \left|\Delta m_{4l}^2\right| < 10^{-3}$ eV$^2$. This allows us to avoid fast oscillations in the T2K/NOvA far detectors and any statistical pathologies that may arise. We find that there remains a preference for four neutrinos over three neutrinos at a level of $\Delta \chi^2 = 4.1$. While this is smaller than what we observed for $\left|\Delta m_{4l}^2\right| \approx 10^{-2}$ eV$^2$, it is nevertheless comparable to the preference for non-standard interactions as a solution to this tension found in Refs.~\cite{Denton:2020uda,Chatterjee:2020kkm} at the level of $\Delta \chi^2 \approx 4.4-4.5$.

\begin{figure}
\begin{center}
\includegraphics[width=0.86\linewidth]{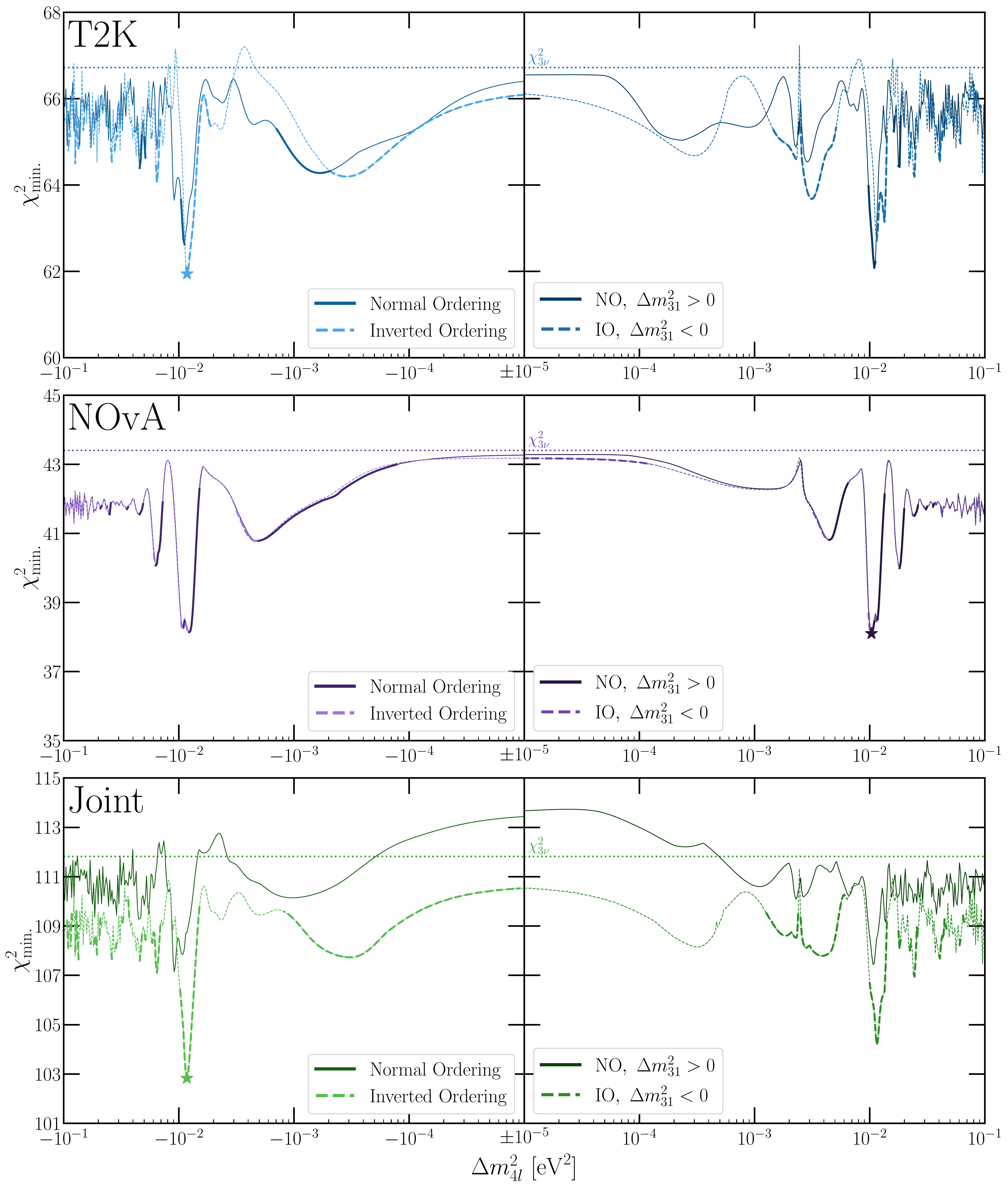}
\caption{Best-fit $\chi^2$ obtained using our analysis of T2K (top, blue), NOvA (middle, purple), and a joint fit of the two (bottom, green) as a function of different values of $\Delta m_{4l}^2$. Different tones within each panel indicate different mass orderings (the signs of $\Delta m_{31}^2$ and $\Delta m_{4l}^2$). The minimization has been performed across all other oscillation parameters except for $\theta_{12}$ and $\Delta m_{21}^2$, which are fixed. \label{fig:Dm41BFP}}
\end{center}
\end{figure}
We generalize this best-fit procedure by, instead of minimizing over all parameters (including $\Delta m_{4l}^2$), scanning over $\Delta m_{4l}^2$ values. We again allow for both positive and negative values of this new mass-squared difference and for both the normal and inverted mass orderings for the three mostly active states. Fig.~\ref{fig:Dm41BFP} presents the results of this approach. The top panels (blue lines) show the results for T2K, middle panels (purple) for NOvA, and bottom panels (green) for the combined analysis. In each row, the left (right) panel corresponds to negative (positive) values of $\Delta m_{4l}^2$. Dark (light) lines in each case correspond to the NO (IO) among the mostly-active neutrinos. Dashed lines in each panel indicate the best-fit $\chi^2$ under the three-neutrino hypothesis presented in Table~\ref{tab:BFPs3nu}. Stars indicate the overall best-fit point of each analysis (when considering all different mass orderings), and lines are made bold if they constitute the minimum $\chi^2$ for a given experimental analysis for all of these choices of mass orderings.

The findings of Table~\ref{tab:BFPs4nuSmall} (and the corresponding tables in Appendix~\ref{app:DetailFit}) are borne out in Fig.~\ref{fig:Dm41BFP}, showing that the fits prefer $\left|\Delta m_{4l}^2\right| \sim 10^{-2}$ eV$^2$ in all cases, with moderate improvements relative to the three-neutrino fits. Above, we discussed the possibility that this preference has to do with the energy resolution and binning of the experiments and the statistical significance when interpreting confidence levels from $\Delta \chi^2$ may be overstated. If we restrict ourselves to $\left|\Delta m_{4l}^2\right| \lesssim 10^{-3}$ eV$^2$ to avoid this concern, we still find moderate preference for a fourth neutrino -- see Appendix~\ref{app:LowDm41} for further discussion.

\begin{figure}
\begin{center}
\includegraphics[width=0.55\linewidth]{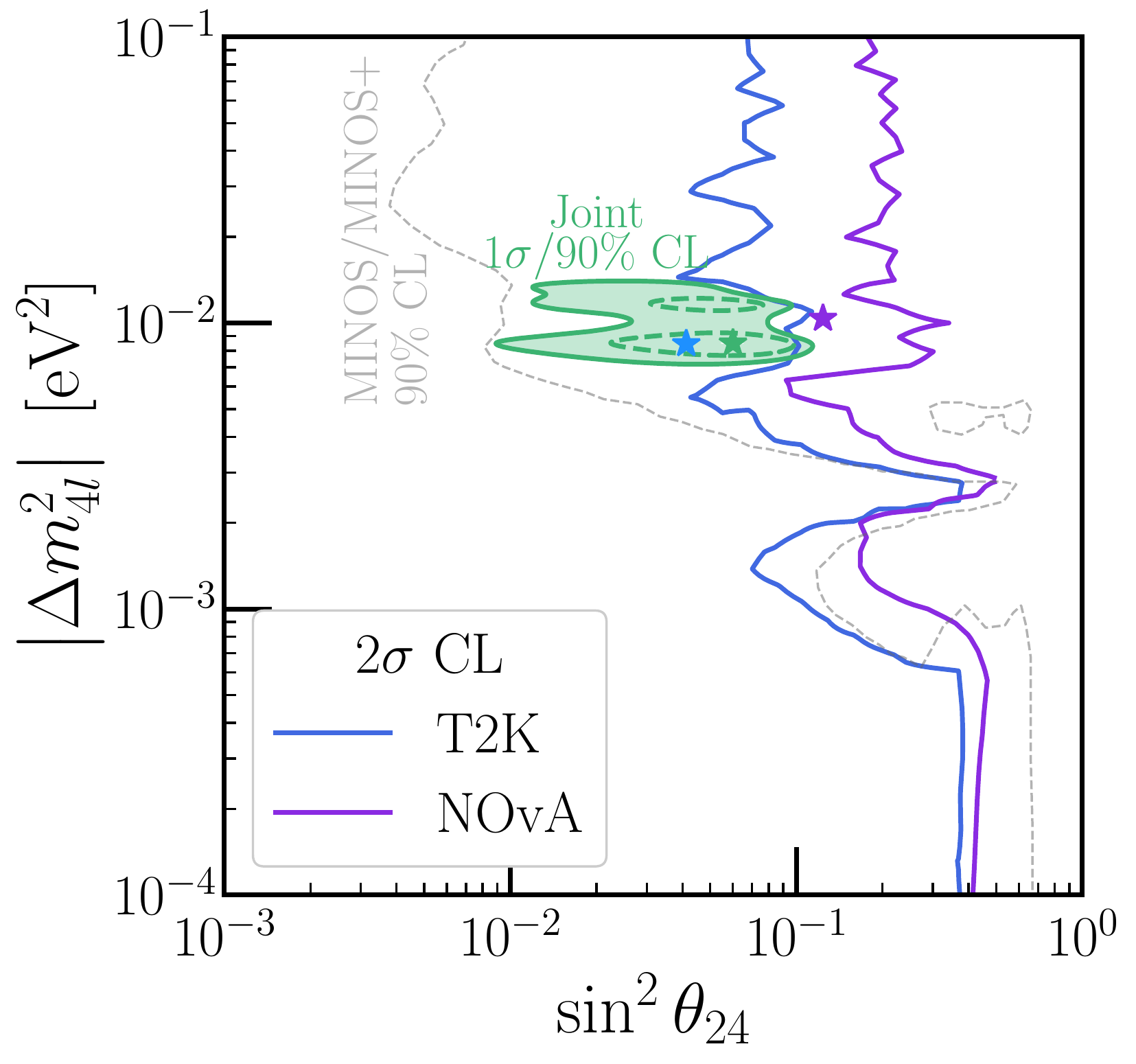}
\caption{Constraints on $\sin^2\theta_{24}$ vs. $\Delta m_{4l}^2$ at $2\sigma$ CL from T2K (blue) and NOvA (purple) after marginalizing over all other parameters (except for $|U_{e2}|^2$ and $\Delta m_{21}^2$, which are fixed and a prior from Daya Bay on $|U_{e3}|^2$ -- see text), including the signs of $\Delta m_{31}^2 $ and $\Delta m_{4l}^2$. The green region indicates the preferred region from a combined analysis at $1\sigma$ (dashed) and 90\% (solid) CL, and the grey, dashed line shows the 90\% CL constraint from MINOS/MINOS+~\cite{MINOS:2017cae}. All confidence levels presented here are derived assuming Wilks' theorem holds.\label{fig:T24Dm41}}
\end{center}
\end{figure}

Moving on from best-fit determinations, we now construct constraints on the new parameters, specifically $\sin^2\theta_{24}$ and $\Delta m_{4l}^2$ (the ones to which these experiments have the greatest sensitivity). In order to present constraints at a particular confidence level and compare against other literature results, we assume for this exercise that Wilks' theorem holds~\cite{Wilks:1938dza}.  After marginalizing over the remaining oscillation parameters (still fixing $|U_{e2}|^2$ and $\Delta m_{21}^2$), we present $2\sigma$ CL constraints from T2K (blue) and NOvA (purple) in Fig.~\ref{fig:T24Dm41}. In generating these constraints, we have marginalized over the signs of both $\Delta m_{31}^2$ and $\Delta m_{4l}^2$. Colored stars indicate the best-fit point in $(\sin^2\theta_{24}, \left|\Delta m_{4l}^2\right|)$ of the given fits.
In Fig.~\ref{fig:T24Dm41} we also compare against the 90\% CL constraint from the MINOS/MINOS+ experiment~\cite{MINOS:2017cae} as a faint grey line.\footnote{This result assumed $\Delta m_{31}^2$ and $\Delta m_{41}^2$ to both be positive, however, due to the lack of mass-ordering sensitivity at MINOS, the result likely does not depend strongly on this choice.} Finally, we also present in green the preferred region at $1\sigma$/90\% CL\footnote{We choose 90\% CL for clarity (the $2\sigma$ CL region spans the entire range of $\left|\Delta m_{4l}^2\right|$ of the figure and a comparable region of $\sin^2\theta_{24}$) and for a direct comparison against the MINOS/MINOS+ result.} ($\Delta \chi^2 = 2.3,\ 4.61$ assuming Wilks' theorem for two parameters) by our combined T2K and NOvA analysis. This result is in tension with that of the MINOS/MINOS+ result, however, our preferred region has not been Feldman-Cousins corrected, and the results would likely agree if a higher confidence level were assumed. T2K has reported constraints in the $\sin^2\theta_{24}$ vs. $\Delta m_{41}^2$ parameter space in Ref.~\cite{T2K:2019efw} -- we find comparable results here despite the simplified assumptions we have made in our analysis and the slightly larger data set considered in this work.

While Fig.~\ref{fig:T24Dm41} compares constraints and preferred regions in the parameter space $\sin^2\theta_{24}$ vs. $\left|\Delta m_{4l}^2\right|$, it is also important to consider the parameters that have been marginalized in this construction. For concreteness, we focus on the preferred region (green) from the combined T2K/NOvA analysis that we have performed. The best-fit point, at $\left|\Delta m_{4l}^2\right| = 8.5 \times 10^{-3}$ eV$^2$, corresponds to mixing angles
\begin{equation}\label{eq:BFAngles}
\left\lbrace \sin^2\theta_{14},\   \sin^2\theta_{24},\   \sin^2\theta_{34} \right\rbrace = \left\lbrace 4.3 \times 10^{-2},\ 6.0\times 10^{-2},\ 0.37\right\rbrace,
\end{equation}
or  mixing-matrix elements
\begin{equation}\label{eq:BFUSqs}
\left\lbrace \left|U_{e4}\right|^2,\ \left|U_{\mu4}\right|^2,\ \left|U_{\tau4}\right|^2\right\rbrace = \left\lbrace 4.3 \times 10^{-2},\ 5.7\times 10^{-2},\ 0.33\right\rbrace.
\end{equation}
For these low values of $\left|\Delta m_{4l}^2\right|$, the strongest constraints on $\left|U_{e4}\right|^2$ come from reactor antineutrino oscillation experiments such as Daya Bay~\cite{DayaBay:2016qvc} and Bugey-3~\cite{Declais:1994su}. A combined analysis~\cite{MINOS:2020iqj} constrains $\sin^2\theta_{14} \lesssim 4 \times 10^{-3}$ at 90\% CL, in significant tension with the value found in Eq.~\eqref{eq:BFAngles}. 

Constraints on $\left|U_{\tau4}\right|^2$ are more difficult to extract, as they often arise in tandem with $\left|U_{\mu4}\right|^2$ and depend strongly on $\Delta m_{41}^2$~\cite{Dentler:2018sju}. While specific constraints in this region of $\left|\Delta m_{4l}^2\right|$ have not been explicitly derived, $\left|U_{\tau4}\right|^2 = 0.33$ is possibly in tension with existing results from neutrino experiments. T2K, which analyzed its neutral-current data in addition to the data sets considered here, has constrained $\left|U_{\tau4}\right|^2 \lesssim 0.5$ for both $\Delta m_{41}^2 = 3 \times 10^{-3}$ eV$^2$ and $0.1$ eV$^2$ at 90\% CL \cite{T2K:2019efw}. Atmospheric neutrino experiments, including Super-Kamiokande~\cite{Super-Kamiokande:2014ndf} and IceCube~\cite{IceCube:2017ivd} have constrained $\left|U_{\tau4}\right|^2 \lesssim 0.2$ at high confidence, however, these analyses are restricted to $\Delta m_{41}^2 \gtrsim 0.1$ eV$^2$ where the fourth-neutrino-driven oscillations are averaged out. A more thorough investigation of this $10^{-2}$ eV$^2$ regime would prove useful if this hint persists in future NOvA/T2K data.

When discussing Fig.~\ref{fig:Dm41BFP}, we considered the possibility of analyzing only the region $\left|\Delta m_{4l}^2\right| \lesssim 10^{-3}$ eV$^2$, in part to avoid concerns regarding energy resolution and bin widths. We noted that in that region, a solution to the NOvA/T2K tension persists with a preference of $\Delta \chi^2 \approx 4.1$. This regime has the added benefit that constraints from MINOS/MINOS+ (as seen in Fig.~\ref{fig:T24Dm41}), Daya Bay/Bugey-3/others, and Super-Kamiokande/IceCube are considerably weaker. Such an \textit{extremely-light} sterile neutrino, as we discuss in Appendix~\ref{app:LowDm41}, with $\left|\Delta m_{4l}^2\right| \approx 7 \times 10^{-4}$ eV$^2$ should be paid particular attention as more data from T2K and NOvA are unveiled, especially if any tension between the two persists.

T2K and NOvA will continue collecting data -- if a very light sterile neutrino does in fact exist with $\left|\Delta m_{4l}^2\right| \approx 10^{-2}$ eV$^2$, more data will continue to shed light and potentially lead to a discovery. In the next generation, the Deep Underground Neutrino Experiment (DUNE)~\cite{DUNE:2020ypp} and Hyper-Kamiokande (HK)~\cite{Hyper-Kamiokande:2018ofw} experiments will have sensitivity to light sterile neutrinos in the same region of $\left|\Delta m_{4l}\right|^2$ given that they operate in a similar $L/E_\nu$ as NOvA and T2K. The two experiments, and any combined analysis, will have excellent sensitivity to test this solution to the T2K/NOvA tension~\cite{Berryman:2015nua,Kelly:2017kch}.

\section{Concluding Remarks}
\label{sec:conclusion}
\setcounter{equation}{0}
As more data from neutrino oscillation experiments are collected, we are able to test the standard-three-massive-neutrinos paradigm with better precision. Concurrently, there is always the possibility that disagreements arise, especially when data from multiple experiments are analyzed. In these instances, exploring different explanations of such tensions is invaluable, whether they are related to statistical fluctuations, deeper systematic issues, or new physics beyond the standard-three-massive-neutrinos paradigm.

Such a tension has been noted when comparing the latest data from the Tokai to Kamioka (T2K) and NuMI Off-axis $\nu_e$ Appearance (NOvA) experiments. These measure the (dis)appearance of $\nu_e$ ($\nu_\mu$) in a $\nu_\mu$ beam at relatively long baselines. When analyzed under the three-neutrino hypothesis, their results disagree at around the 90\% confidence level. Previous studies of combination T2K and NOvA data have highlighted that this tension is reduced when, for instance, the inverted neutrino mass ordering is considered instead of the normal ordering~\cite{Kelly:2020fkv,Esteban:2020cvm,deSalas:2020pgw,Capozzi:2021fjo}, or when additional, beyond-the-Standard-Model neutrino/matter interactions are included in the analyses~\cite{Denton:2020uda,Chatterjee:2020kkm}.

We have demonstrated here that an alternative approach can remedy this tension -- the addition of a fourth, very light, sterile neutrino. This very light new neutrino would be associated to a mass-squared difference, relative to the lightest mostly-active neutrino, of order $10^{-2}$ eV$^2$. We have studied the four-neutrino hypothesis when applied to the T2K and NOvA data independently, as well as their combination. For the combined data, we find that the four-neutrino hypothesis is preferred over the three-neutrino one at the level of $\Delta \chi^2 \approx 9$. When interpreting this in terms of statistical significance, two difficulties arise: first, the number of additional parameters in the four-neutrino hypothesis relative to the three-neutrino one (six additional parameters). Second, the oscillations associated with a new mass-squared difference on the order of $10^{-2}$ eV$^2$ are significant within individual bins in these long-baseline experiments, which leads to an artificial preference for sterile neutrinos due to statistical fluctuations.

Due to the second challenge, in order to avoid relatively fast oscillations, we also explored an alternative extremely-light sterile neutrino analysis where the fourth neutrino is fixed to be associated to a mass-squared difference smaller (in magnitude) than $10^{-3}$ eV$^2$. In this context, we find moderate improvement relative to the three-neutrino hypothesis, at the level of $\Delta \chi^2 \approx 4$. While this is less significant, it is comparable to the improvement offered by non-standard neutrino interactions and merits further investigation.

NOvA and T2K are still collecting and analyzing data. As they progress, the experiments and combined analyses thereof will allow for deeper testing of these different, interesting regimes of four-neutrino oscillations with a very light or extremely light fourth neutrino. If they confirm the existence of such a new, light fermion state, then future experiments (including the spiritual successors DUNE and Hyper-Kamiokande) will be able to probe the new particle's properties with even greater precision.

\section*{Acknowledgements}
This work was supported in part by the US Department of Energy (DOE) grant \#de-sc0010143 and in part by the National Science Foundation under Grant Nos.~PHY-1630782 and PHY-1748958. The document was prepared using the resources of the Fermi National Accelerator Laboratory (Fermilab), a DOE, Office of Science, HEP User Facility. Fermilab is managed by Fermi Research Alliance, LLC (FRA), acting under Contract No. DE-AC02-07CH11359.

\appendix
\section{Detailed Fit Results}\label{app:DetailFit}

In Section~\ref{sec:Results}, we provided best-fit points of our analyses of T2K, NOvA, and their combination under the three- and four-neutrino hypotheses. When discussing the best-fit points under the four-neutrino hypothesis (Table~\ref{tab:BFPs4nuSmall}), we showed the results of the analysis (i.e. which signs of $\Delta m_{31}^2$ and $\Delta m_{4l}^2$) that provided the best overall fit to each experimental data set. In this appendix, we provide the results to all four fits for each experiment/combination. Table~\ref{tab:BFPs4nuT2KNOvA} does so for our analyses of T2K and NOvA data separately, and Table~\ref{tab:BFPs4nuJoint} does so for their combination.

\begin{table}
\begin{center}
\caption{Best-fit 4$\nu$ parameters of our four T2K (left) and NOvA (right) analyses. See Section~\ref{subsec:4nuResults} for more detail.\label{tab:BFPs4nuT2KNOvA}}
{\footnotesize
\begin{tabular}{c||c|c|c|c|}
\multirow{3}{*}{{\large $4\nu$}} & \multicolumn{4}{c|}{{\large T2K}} \\ \cline{2-5}
& \multicolumn{2}{c|}{NO} & \multicolumn{2}{c|}{IO} \\ \cline{2-5}
& $\Delta m_{4l}^2 > 0$ & $<0$ & $>0$ & $<0$ \\ \hline \hline
$\sin^2\theta_{13}$ & $0.024$ & $0.024$ & $0.024$ & $0.024$  \\ \hline
$\sin^2\theta_{23}$ & $0.44$ & $0.44$ & $0.44$ & $0.43$  \\ \hline
$\Delta m_{31}^2/10^{-3}$ eV$^2$ & $2.49$ & $2.48$ & $-2.38$ & $-2.39$ \\ \hline
$\delta_{\rm CP}$ & $4.94$ & $4.89$ & $4.45$ & $4.42$  \\ \hline\hline
$\sin^2\theta_{14}$ & $7.1\times 10^{-2}$ & $7.8\times 10^{-2}$ & $8.0\times 10^{-2}$ & $7.8\times 10^{-2}$  \\ \hline
$\sin^2\theta_{24}$ & $4.2\times 10^{-2}$ & $4.0\times 10^{-2}$ & $4.1\times 10^{-2}$ & $4.1\times 10^{-2}$  \\ \hline
$\sin^2\theta_{34}$ & $5.2\times 10^{-2}$ & $5.2\times 10^{-2}$ & $5.6\times 10^{-1}$ & $7.8\times 10^{-1}$  \\ \hline
$\Delta m_{4l}^2/$eV$^2$ & $1.1\times 10^{-2}$ & $-9.0\times 10^{-3}$ & $1.1\times 10^{-2}$ & $-8.5\times 10^{-3}$ \\ \hline
$\delta_{14}$ & $3.51$ & $3.14$ & $2.08$ & $1.83$  \\ \hline
$\delta_{24}$ & $6.10$ & $5.89$ & $2.72$ & $2.64$  \\ \hline\hline
$\chi^2_{4\nu}$ & $62.07$ & $62.63$ & $62.80$ & $61.95$  \\ \hline
Best-fit & \multicolumn{4}{c|}{$m_4 < m_3 < m_1 < m_2$} \\ \hline
$\chi^2_{3\nu} - \chi^2_{4\nu}$  & \multicolumn{4}{c|}{$4.87$}
\end{tabular}}
\quad
{\footnotesize
\begin{tabular}{|c|c|c|c|}
\multicolumn{4}{|c|}{{\large NOvA}} \\ \hline
\multicolumn{2}{|c|}{NO} & \multicolumn{2}{c|}{IO} \\ \hline
$\Delta m_{4l}^2 > 0$ & $<0$ & $>0$ & $<0$ \\ \hline \hline
$0.022$ & $0.022$ & $0.022$ & $0.022$  \\ \hline
$0.44$ & $0.62$ & $0.59$ & $0.41$  \\ \hline
$2.43$ & $2.44$ & $-2.32$ & $-2.35$ \\ \hline
$0.00$ & $5.22$ & $3.19$ & $4.58$  \\ \hline\hline
$6.9\times 10^{-3}$ & $1.6\times 10^{-2}$ & $8.9\times 10^{-3}$ & $1.4\times 10^{-2}$  \\ \hline
$1.2\times 10^{-1}$ & $1.2\times 10^{-1}$ & $1.3\times 10^{-1}$ & $1.1\times 10^{-1}$  \\ \hline
$0.29$ & $0.79$ & $0.34$ & $0.69$  \\ \hline
$1.0\times 10^{-2}$ & $-8.0\times10^{-3}$ & $1.0\times10^{-2}$ & $-8.1\times10^{-3}$ \\ \hline
$3.51$ & $4.07$ & $4.81$ & $4.69$ \\ \hline
$3.15$ & $3.21$ & $0.12$ & $0.15$  \\ \hline\hline
$38.10$ & $38.14$ & $38.13$ & $38.16$  \\ \hline
\multicolumn{4}{|c|}{$m_1 < m_2 < m_3 < m_4$} \\ \hline
\multicolumn{4}{|c|}{$5.30$}
\end{tabular}}

\end{center}
\end{table}

\begin{table}
\begin{center}
\caption{Best-fit 4$\nu$ parameters of our four combined T2K+NOvA analyses. See Section~\ref{subsec:4nuResults} for more detail.\label{tab:BFPs4nuJoint}}
{\scriptsize
\begin{tabular}{c||c|c|c|c|}
\multirow{3}{*}{{\large $4\nu$}} & \multicolumn{4}{c|}{{\large Combined T2K and NOvA}} \\ \cline{2-5}
& \multicolumn{2}{c|}{NO} & \multicolumn{2}{c|}{IO} \\ \cline{2-5}
& $\Delta m_{4l}^2 > 0$ & $<0$ & $>0$ & $<0$ \\ \hline \hline
$\sin^2\theta_{13}$ & $0.023$ & $0.025$ & $0.023$ & $0.023$  \\ \hline
$\sin^2\theta_{23}$ & $0.45$ & $0.45$ & $0.44$ & $0.43$  \\ \hline
$\Delta m_{31}^2/10^{-3}$ eV$^2$ & $2.49$ & $2.51$ & $-2.36$ & $-2.39$ \\ \hline
$\delta_{\rm CP}$ & $4.09$ & $3.88$ & $1.72$ & $4.47$ \\ \hline\hline
$\sin^2\theta_{14}$ & $2.1\times 10^{-2}$ & $1.1\times 10^{-1}$ & $3.4\times 10^{-2}$ & $4.3\times 10^{-2}$ \\ \hline
$\sin^2\theta_{24}$ & $5.3\times 10^{-2}$ & $3.3\times 10^{-2}$ & $5.3\times 10^{-2}$ & $6.0\times 10^{-2}$ \\ \hline
$\sin^2\theta_{34}$ & $0.56$ & $0.21$ & $1.1\times 10^{-2}$ & $0.37$ \\ \hline
$\Delta m_{4l}^2/$eV$^2$ & $1.1\times10^{-2}$ & $-1.1\times10^{-2}$ & $1.2\times10^{-2}$ & $-8.5\times 10^{-3}$ \\ \hline
$\delta_{14}$ & $0.01$ & $0.03$ & $6.09$ & $4.88$ \\ \hline
$\delta_{24}$ & $1.82$ & $1.18$ & $0.53$ & $5.89$  \\ \hline\hline
$\chi^2_{4\nu}$ &$107.41$ & $107.62$ & $104.27$ & $102.83$  \\ \hline
Best-fit & \multicolumn{4}{c|}{$m_4 < m_3 < m_1 < m_2$} \\ \hline
$\chi^2_{3\nu} - \chi^2_{4\nu}$  & \multicolumn{4}{c|}{$8.99$}
\end{tabular}}
\end{center}
\end{table}

\section{Alternative Analyses with Very Small Mass-Squared Difference}\label{app:LowDm41}

We find, in Section~\ref{sec:Results}, a solution to the NOvA/T2K tension with a new, light sterile neutrino with a mass-squared difference $\left|\Delta m_{4l}^2\right| \approx 10^{-2}$ eV$^2$. However, there are technical challenges associated with this relatively large mass-squared difference for the NOvA/T2K analyses, also as discussed in Section~\ref{sec:Results}. For those reasons, we choose to pursue a different version of the analyses from the main text, this time restricting ourselves to $\left|\Delta m_{4l}^2\right| \leq 10^{-3}$ eV$^2$. As with the analyses in the main text, we fix $\Delta m_{21}^2$ to its best-fit value ($7.53\times 10^{-5}$ eV$^2$).

\begin{figure}[htbp]
\begin{center}
\includegraphics[width=0.9\linewidth]{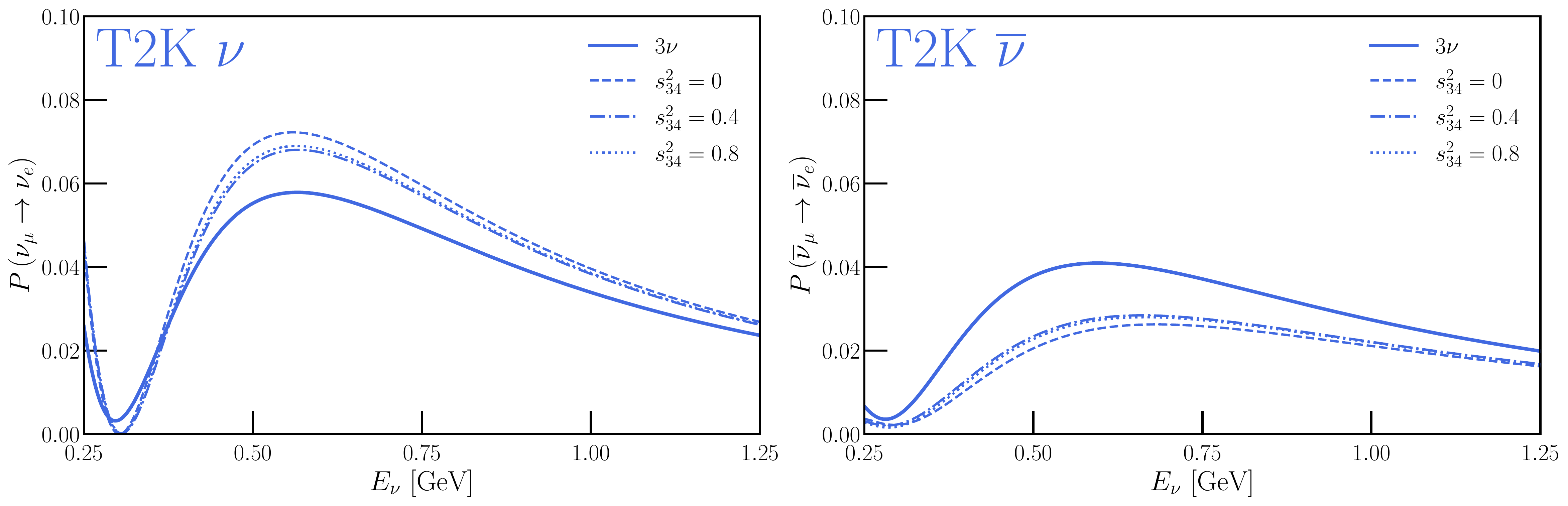}
\includegraphics[width=0.9\linewidth]{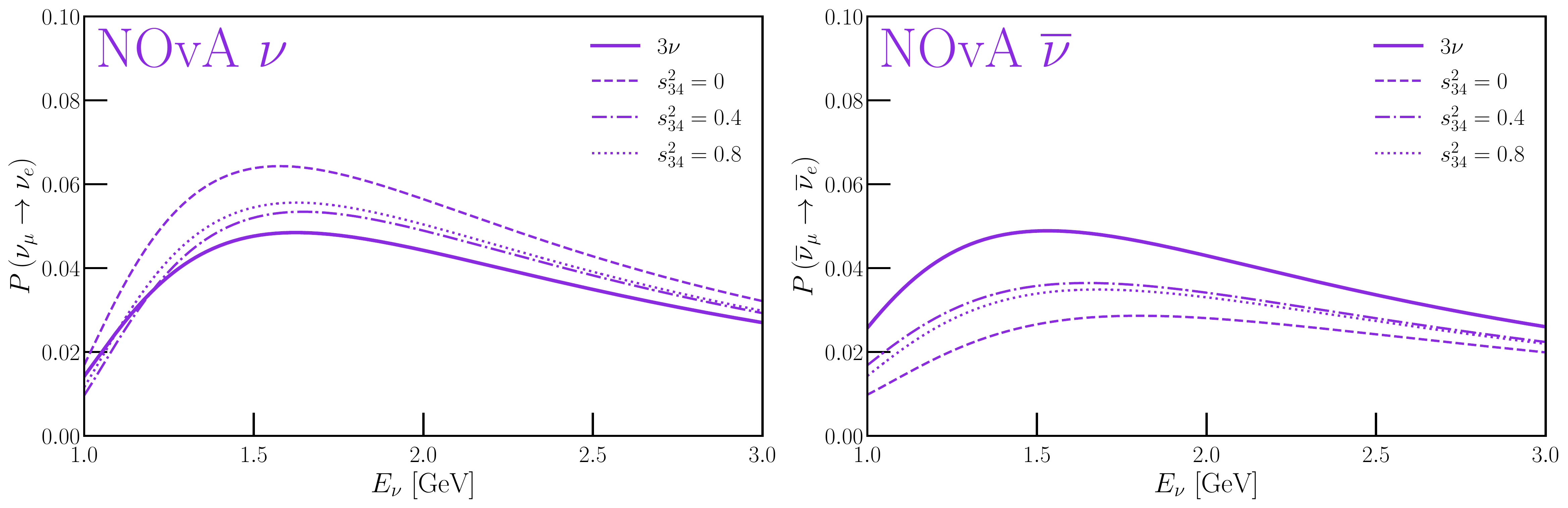}
\caption{Oscillation probabilities at T2K (top) and NOvA (bottom) comparing three-neutrino oscillation probabilities (solid lines, parameters from Table~\ref{tab:OscParams}) against four-neutrino ones (non-solid lines, parameters from the ``Joint'' column in Table~\ref{tab:BFPs4nuSmall_LowDm41}). Left panels show probabilities for neutrino oscillation, whereas right ones show antineutrino oscillation. For the four-neutrino probabilities, three choices of $\sin^2\theta_{34}$ are used for demonstration: dashed/dot-dashed/dotted lines correspond to $\sin^2\theta_{34} = 0,\ 0.4,\ 0.8$. \label{fig:OscProbsLow}}
\end{center}
\end{figure}

First, we illustrate how the oscillation probabilities $P(\nu_\mu \to \nu_e)$ and $P(\overline\nu_\mu \to \overline\nu_e)$ at T2K/NOvA energies and baselines behave for a very light sterile neutrino, similar to the discussion in Section~\ref{sec:FourFlavor} (see Fig.~\ref{fig:plainprobs}). Instead of a relatively large $\left|\Delta m_{4l}^2\right| \approx 10^{-2}$ eV$^2$, Fig.~\ref{fig:OscProbsLow} depicts the impact of a new mass-squared difference $\Delta m_{4l}^2 = -3.4 \times 10^{-4}$ eV$^2$ (and an inverted mass ordering for the three mostly-active neutrinos). The remaining oscillation parameters we use are from the ``Joint'' column in Table~\ref{tab:BFPs4nuSmall_LowDm41}, corresponding to the best-fit parameters of the combined T2K and NOvA analysis when the new mass-squared difference is restricted to be $\left|\Delta m_{4l}^2\right| \leq 10^{-3}$ eV$^2$.

\begin{figure}[htbp]
\begin{center}
\includegraphics[width=0.9\linewidth]{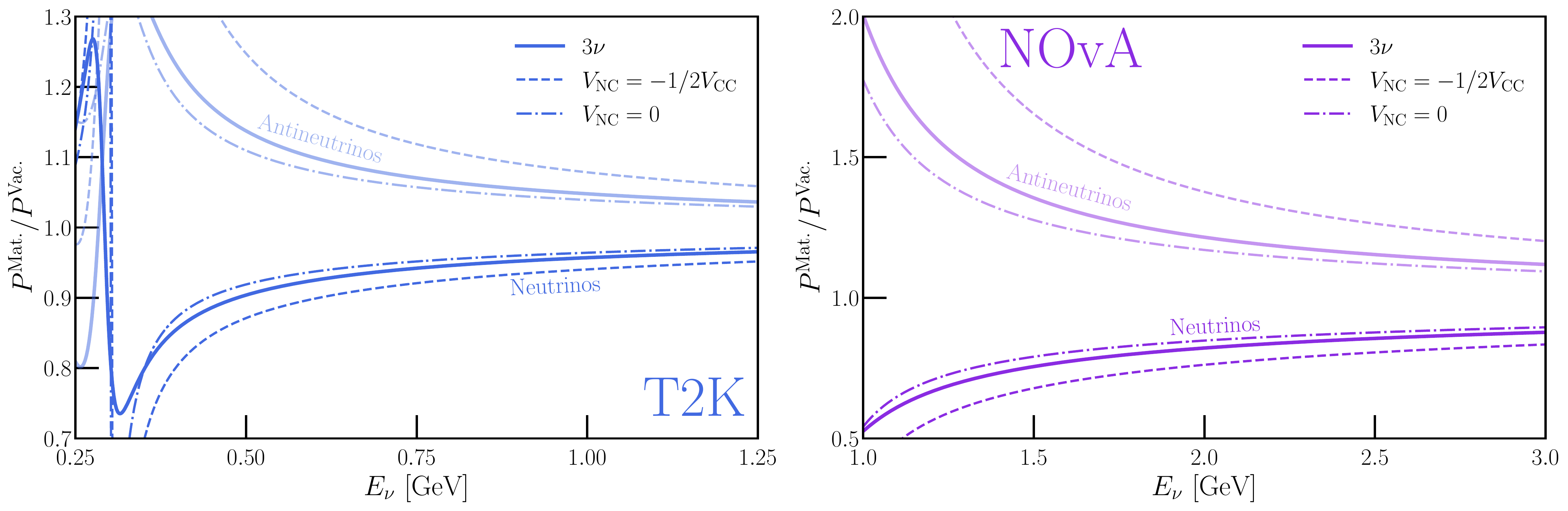}
\caption{Ratio of oscillation probabilities, similar to Fig.~\ref{fig:ratioprobs}, considering an extremely light sterile neutrino with $\Delta m_{4l}^2 = -3.4 \times 10^{-4}$ eV$^2$ and oscillation parameters as given in Table~\ref{tab:BFPs4nuSmall_LowDm41}. \label{fig:RatioLow}}
\end{center}
\end{figure}

The top panels of Fig.~\ref{fig:OscProbsLow} show oscillation probabilities at T2K, and the bottom panels at NOvA; the left (right) panels correspond to neutrino (antineutrino) oscillations. As with Fig.~\ref{fig:plainprobs}, we allow $\sin^2\theta_{34}$ to vary to demonstrate its nontrivial impact on these oscillation probabilities -- the dashed/dot-dashed/dotted lines correspond to $\sin^2\theta_{34} = 0,\ 0.4,\ 0.8$, respectively. Compared with Fig.~\ref{fig:plainprobs}, here the ``new'' oscillation length driven by $\Delta m_{21}^2 < \left|\Delta m_{4l}^2\right| < \left|\Delta m_{31}^2\right|$ is relatively long as a function of the neutrino energy, leading at zeroth order to an overall shift in normalization relative to the three-neutrino oscillation probabilities. Across the energies of interest for T2K and NOvA, this leads to larger values of $P(\nu_\mu \to \nu_e)$ and smaller values of $P(\overline\nu_\mu \to \overline\nu_e)$. As in Fig.~\ref{fig:plainprobs}, the impact of nonzero $\sin^2\theta_{34}$ is more prevalent for NOvA, with its longer baseline, than for T2K. Fig.~\ref{fig:RatioLow} depicts the impact of matter effects for this relatively smaller value of $\Delta m^2_{4l}$ and is to be compared to Fig.~\ref{fig:ratioprobs}.

The best-fit points obtained from this low-$\Delta m^2_{4l}$ fit to T2K data, NOvA data, and the combined data sets are listed in Table~\ref{tab:BFPs4nuSmall_LowDm41}. As in the result discussed in the main text, NOvA favors NO for the mostly active states while T2K and the Joint fits favor the IO for the mostly active states. All fits point to $m_4$ as the lightest neutrino mass. The improvement relative to the three-neutrinos scenario is largest for the Joint fit -- a little over four units of $\chi^2$ -- but rather modest. In summary, the data do not significantly favor the four-neutrino hypothesis over the three-neutrino one.  

\begin{figure}
\begin{center}
\includegraphics[width=0.55\linewidth]{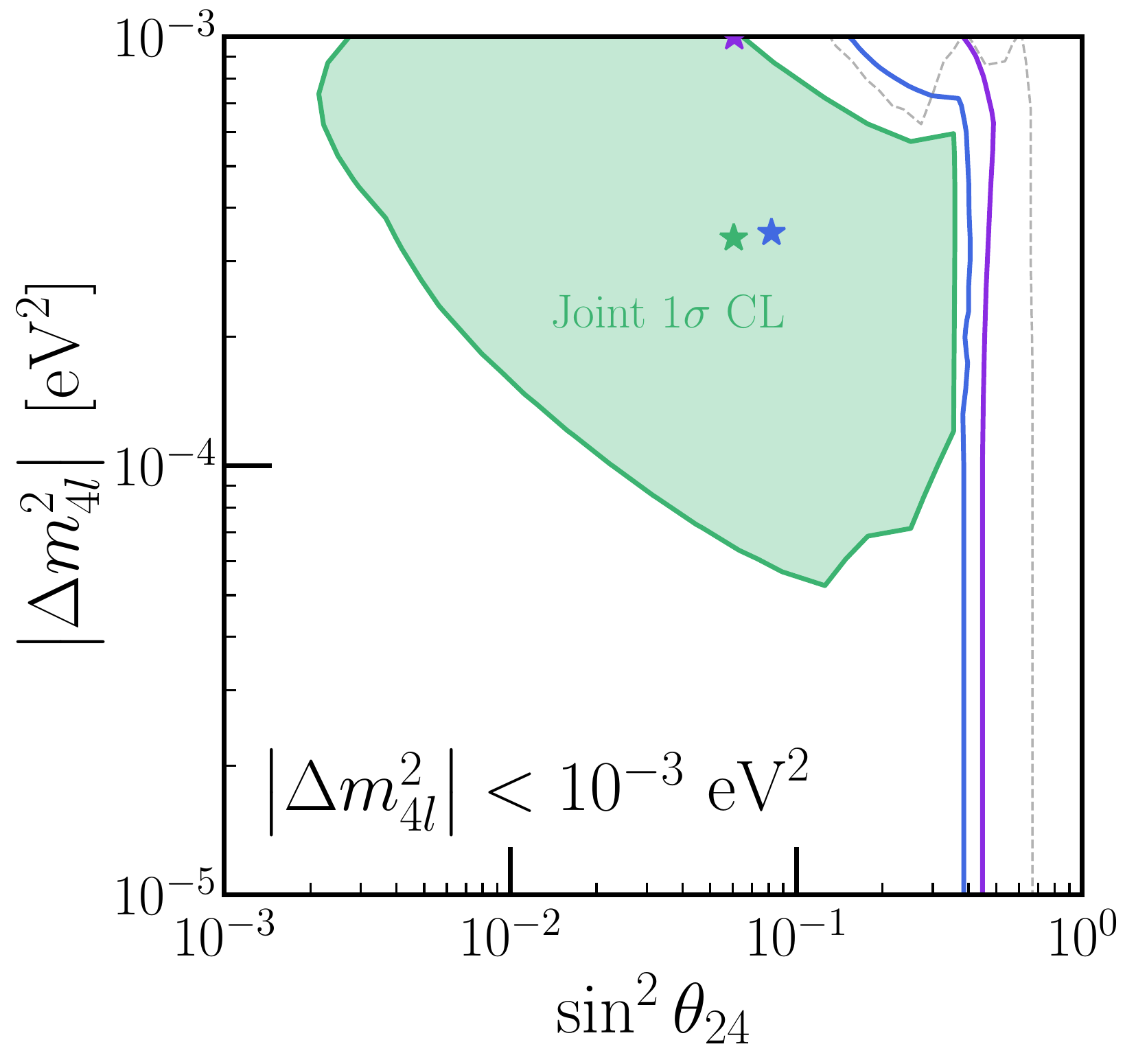}
\caption{Similar to Fig.~\ref{fig:T24Dm41} but under the analysis assumption that $\left| \Delta m_{4l}^2\right| < 10^{-3}$ eV$^2$.\label{fig:T24Dm41Low}}
\end{center}
\end{figure}

Fig.~\ref{fig:T24Dm41Low} depicts the region of the $\left|\Delta m_{4l}^2\right|\times\sin^2\theta_{24}$ parameter space that is allowed by the combination of T2K and NOvA data at the one-sigma level, including all possible four-neutrino mass orderings (see Fig.~\ref{fig:MO}) and assuming $\left|\Delta m^2_{4l}\right|$ is less than $10^{-3}$~eV$^2$, along with the 2$\sigma$ constraints from NOvA (purple) and T2K (blue). The stars indicate the best-fit points and the dashed line existing bounds from MINOS/MINOS+. Unlike the result discussed in the main text, here the best fit point is not in tension with existing neutrino oscillation bounds thanks to the more limited sensitivity of MINOS/MINOS+ and reactor antineutrino experiments to new mass-squared differences less than $10^{-3}$~eV$^2$. 

Like the results discussed in the main text, here, the best-fit points in Table~\ref{tab:BFPs4nuSmall_LowDm41} all prefer large values of $\sin^2\theta_{34}$, i.e., they suggest that $\nu_4$ has an ${\cal O}(1)$  $\nu_{\tau}$ component. As discussed in Section~\ref{sec:Results}, while large $\sin^2\theta_{34}$ are excluded by existing data, relevant constraints were obtained only for relatively large $\left|\Delta m^2_{4l}\right|\gtrsim 0.1$~eV$^2$.

\begin{table}[htbp]
\begin{center}
\caption{Best-fit parameters of our $4\nu$ analyses when restricted to $\left|\Delta m_{4l}^2\right| \leq 10^{-3}$ eV$^2$. Other details identical to Table~\ref{tab:BFPs4nuSmall}.\label{tab:BFPs4nuSmall_LowDm41}}
{\footnotesize
\begin{tabular}{c||c|c|c|}
4$\nu$ & T2K & NOvA & Joint \\ \hline \hline
$\sin^2\theta_{13}$ & $0.025$ & $0.022$ & $0.026$ \\ \hline
$\sin^2\theta_{23}$ & $0.41$ & $0.63$ & $0.53$ \\ \hline
$\Delta m_{31}^2/10^{-3}$ eV$^2$ & $-2.37$ & $2.44$ & $-2.39$ \\ \hline
$\delta_{\rm CP}$ & $4.05$ & $2.98$ & $4.21$ \\ \hline\hline
$\sin^2\theta_{14}$ & $0.13$ & $6.2\times 10^{-3}$ & $0.14$ \\ \hline
$\sin^2\theta_{24}$ & $8.2\times 10^{-2}$ & $6.1\times 10^{-2}$ & $7.6\times 10^{-2}$ \\ \hline
$\sin^2\theta_{34}$ & $0.63$ & $0.79$ & $0.48$ \\ \hline
$\Delta m_{4l}^2$/eV$^2$ & $-3.5\times 10^{-4}$ & $-1.0\times 10^{-3}$ & $-3.4\times 10^{-4}$ \\ \hline
$\delta_{14}$ & $4.66$ & $2.77$ & $5.34$ \\ \hline
$\delta_{24}$ & $5.04$ & $3.21$ & $5.39$ \\ \hline\hline
$\chi^2_{\rm 4\nu}$ & $64.20$ & $41.50$ & $5.39$ \\ \hline
Ordering & $m_4 < m_3 < m_1 < m_2$ & $m_4 < m_1 < m_2 < m_3$ & $m_4 < m_3 < m_1 < m_2$ \\ \hline
$\chi^2_{\rm 3\nu} - \chi^2_{\rm 4nu}$ & $2.62$ & $1.90$ & $4.11$ 
\end{tabular}}
\end{center}
\end{table}

\section{Test Statistic Studies and Pseudoexperiments}\label{app:Pseudoexperiments}
Section~\ref{sec:Results} demonstrated that all three fits, those to the T2K and NOvA data individually as well as their combination, prefer the four-neutrino hypothesis over the three-neutrino one to some degree of confidence. This is expected, as the three-neutrino hypothesis is a subset of the four-neutrino one -- what is more difficult to predict is the level at which this preference is found. Specifically, we found that the best-fit-point to the data under the four-neutrino hypothesis compared to that of the three-neutrino hypothesis for T2K, NOvA, and the joint fit exhibited a preference at the level of $\Delta\chi^2 = 4.87,\ 5.30$, and $8.99$, respectively. Also in Section~\ref{sec:Results}, we discussed the fact that these three fits tend to favor $\left|\Delta m_{4l}^2\right| \approx 10^{-2}$ eV$^2$ and opined on whether this is a coincidence due to the binning used by T2K and NOvA or a real, physical effect.

In this appendix, we attempt to quantify some of these observed challenges -- how significant these preferences are, and whether the preferred new mass-squared splitting is spurious. To do so, we perform a number of pseudoexperiments corresponding to each analysis. We simulate data for each experiment assuming the three-neutrino hypothesis is true, assuming $\sin^2\theta_{12} = 0.304$, $\sin^2\theta_{13} = 0.0212$, $\sin^2\theta_{23} = 0.532$, $\Delta m_{21}^2 = 7.53 \times 10^{-5}$ eV$^2$, $\Delta m_{31}^2 = 2.45\times 10^{-3}$ eV$^2$, and $\delta_{\rm CP} = 4.39$ (given as reference values in Ref.~\cite{T2K:2021xwb}). For each pseudoexperiment, we include Poissonian fluctuations on the expected data according to this hypothesis. Then, using the same analysis strategies as in the main text, we obtain the best-fit-points and $\chi^2$ values for the three-neutrino and four-neutrino hypotheses.

\begin{figure}
\begin{center}
\includegraphics[width=0.6\linewidth]{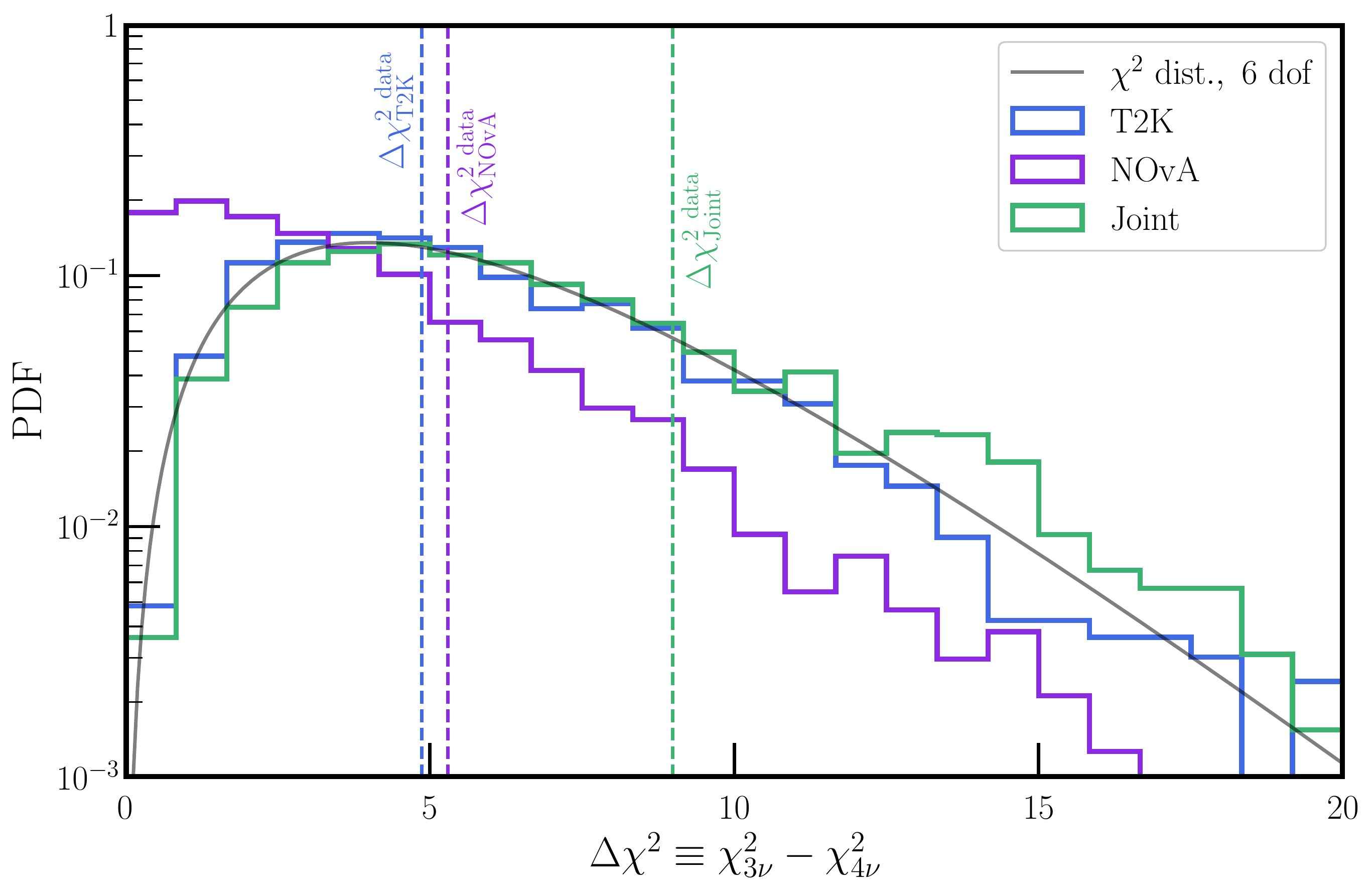}
\caption{Preference for the four-neutrino hypothesis over the three-neutrino one as indicated by pseudoexperiments simulating T2K (blue), NOvA (purple), and their combination (green). We also display the PDF of the chi-squared distribution assuming six degrees of freedom (grey), as well as the preferences indicated when analyzing the actual data sets (dashed lines).\label{fig:PS:Chi2}}
\end{center}
\end{figure}

The normalized distribution of $\Delta\chi^2 \equiv\chi^2_{\rm 3\nu} - \chi^2_{\rm 4\nu}$ is shown in Fig.~\ref{fig:PS:Chi2}. We show the histograms obtained by performing pseudoexperiments of the three different analyses in solid, colored lines, compared against the $\Delta \chi^2$ obtained when analyzing the data as vertical, dashed lines. We also display the $\chi^2$ distribution assuming six degrees of freedom (corresponding to the difference between the number of parameters in the two analysis hypotheses) as a grey line, which seems to track the distribution of the joint-fit pseudoexperiments well.
As a result of this procedure, we can determine the statistical significances of the three preferences -- the p-values of the observed data at T2K, NOvA, and their combination are $0.53$, $0.21$, and $0.22$, respectively. These values correspond to preference for the four-neutrino hypothesis at the level of $0.58\sigma$, $1.26\sigma$, and $1.22\sigma$ -- none of which corresponds to a significant preference.

\begin{figure}
\begin{center}
\includegraphics[width=0.95\linewidth]{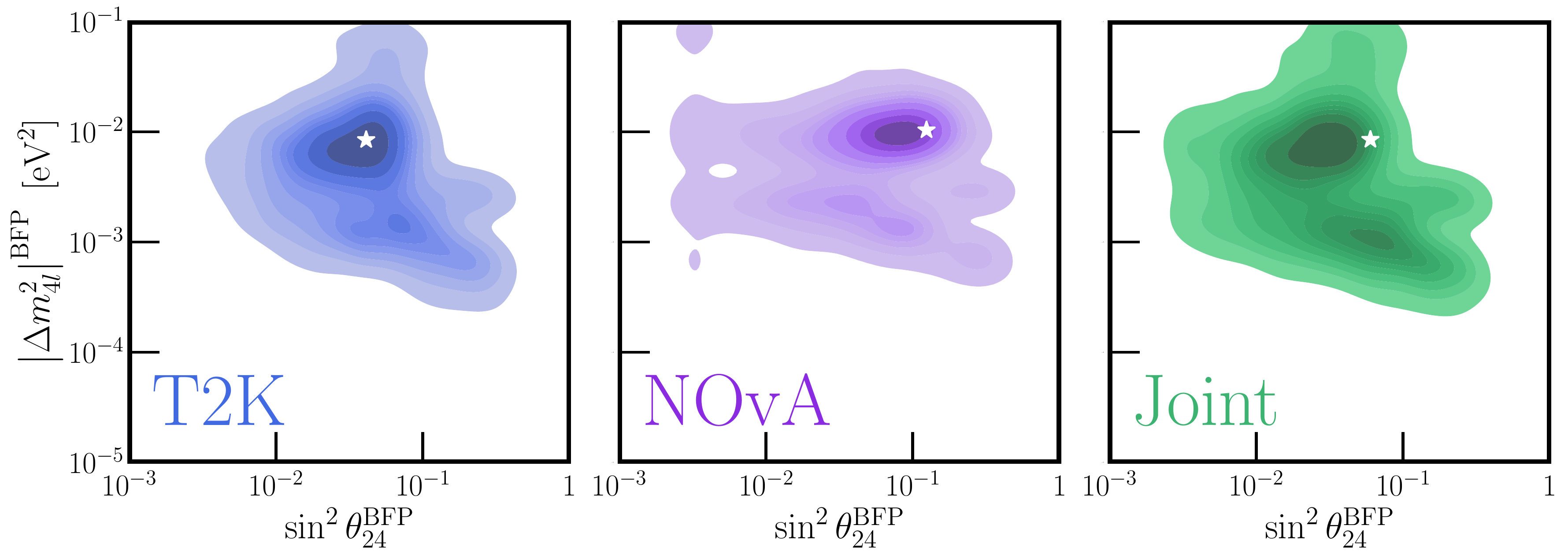}
\caption{Best-fit values of $\sin^2\theta_{24}$ and $\left|\Delta m_{4l}^2\right|$ obtained when performing pseudoexperiments of T2K (left), NOvA (center), and their combination (right). White stars in each panel indicate the best-fit values of these parameters when analyzing the corresponding data set. In the colored distributions, darker (lighter) colors indicate regions where the fit prefers the values more (less) frequently.\label{fig:PS:BFPs}}
\end{center}
\end{figure}

Finally, we determine whether the best-fit points obtained when analyzing data, all with $\left|\Delta m_{4l}^2\right| \approx 10^{-2}$ eV$^2$ are expected when including Poissonian fluctuations of simulated three-neutrino data. We determine, for each pseudoexperiment, the best-fit values of $\sin^2\theta_{24}$ and $\left|\Delta m_{4l}^2\right|$ obtained when analyzing the pseudodata under the four-neutrino hypothesis, displaying the distributions of these best-fit values in Fig.~\ref{fig:PS:BFPs}. Here, the dark regions indicate where the fits prefer the combination of parameters most frequently, and the white stars show the best-fit parameters obtained in each analysis from Section~\ref{sec:Results}.
For all three analyses, the best-fit obtained when analyzing the data is nearly exactly consistent with the most likely points obtained by these procedures. This indicates that such fit values of $\left|\Delta m_{4l}^2\right|$ are to be expected due to the construction of the test statistic and the experimental particulars, furthering the evidence that the results obtained in the main text are due to statistical fluctuations instead of the actual presence of a fourth, very light neutrino.

\bibliographystyle{utphys}
\bibliography{References}

\providecommand{\href}[2]{#2}\begingroup\raggedright\begin{thebibliography}{10}

\bibitem{T2K:2011ypd}
{\bf T2K} , K.~Abe {\em et al.}, ``{Indication of Electron Neutrino Appearance
  from an Accelerator-produced Off-axis Muon Neutrino Beam},''
  \href{http://dx.doi.org/10.1103/PhysRevLett.107.041801}{{\em Phys. Rev.
  Lett.} {\bf 107} (2011)  041801}, \href{http://arxiv.org/abs/1106.2822}{{\tt
  arXiv:1106.2822 [hep-ex]}}.

\bibitem{T2K:2021xwb}
{\bf T2K} , K.~Abe {\em et al.}, ``{Improved constraints on neutrino mixing
  from the T2K experiment with $\mathbf{3.13\times10^{21}}$ protons on
  target},'' \href{http://dx.doi.org/10.1103/PhysRevD.103.112008}{{\em Phys.
  Rev. D} {\bf 103} (2021) no.~11, 112008},
  \href{http://arxiv.org/abs/2101.03779}{{\tt arXiv:2101.03779 [hep-ex]}}.

\bibitem{NOvA:2016kwd}
{\bf NOvA} , P.~Adamson {\em et al.}, ``{First measurement of electron neutrino
  appearance in NOvA},''
  \href{http://dx.doi.org/10.1103/PhysRevLett.116.151806}{{\em Phys. Rev.
  Lett.} {\bf 116} (2016) no.~15, 151806},
  \href{http://arxiv.org/abs/1601.05022}{{\tt arXiv:1601.05022 [hep-ex]}}.

\bibitem{NOvA:2021nfi}
{\bf NOvA} , M.~A. Acero {\em et al.}, ``{An Improved Measurement of Neutrino
  Oscillation Parameters by the NOvA Experiment},''
  \href{http://arxiv.org/abs/2108.08219}{{\tt arXiv:2108.08219 [hep-ex]}}.

\bibitem{T2KNu2020}
P.~Dunne, ``{L}atest {N}eutrino {O}scillation {R}esults from {T2K},
  doi:10.5281/zenodo.3959558,'' July, 2020.
\newblock \url{https://doi.org/10.5281/zenodo.3959558}.

\bibitem{NOvANu2020}
A.~Himmel, ``{N}ew {O}scillation {R}esults from the {NOvA} {E}xperiment,
  doi:10.5281/zenodo.3959581,'' July, 2020.
\newblock \url{https://doi.org/10.5281/zenodo.3959581}.

\bibitem{Kelly:2020fkv}
K.~J. Kelly, P.~A.~N. Machado, S.~J. Parke, Y.~F. Perez-Gonzalez, and R.~Z.
  Funchal, ``{Neutrino mass ordering in light of recent data},''
  \href{http://dx.doi.org/10.1103/PhysRevD.103.013004}{{\em Phys. Rev. D} {\bf
  103} (2021) no.~1, 013004}, \href{http://arxiv.org/abs/2007.08526}{{\tt
  arXiv:2007.08526 [hep-ph]}}.

\bibitem{Esteban:2020cvm}
I.~Esteban, M.~C. Gonzalez-Garcia, M.~Maltoni, T.~Schwetz, and A.~Zhou, ``{The
  fate of hints: updated global analysis of three-flavor neutrino
  oscillations},'' \href{http://dx.doi.org/10.1007/JHEP09(2020)178}{{\em JHEP}
  {\bf 09} (2020)  178}, \href{http://arxiv.org/abs/2007.14792}{{\tt
  arXiv:2007.14792 [hep-ph]}}.

\bibitem{deSalas:2020pgw}
P.~F. de~Salas, D.~V. Forero, S.~Gariazzo, P.~Mart\'\i{}nez-Mirav\'e, O.~Mena,
  C.~A. Ternes, M.~T\'ortola, and J.~W.~F. Valle, ``{2020 global reassessment
  of the neutrino oscillation picture},''
  \href{http://dx.doi.org/10.1007/JHEP02(2021)071}{{\em JHEP} {\bf 02} (2021)
  071}, \href{http://arxiv.org/abs/2006.11237}{{\tt arXiv:2006.11237
  [hep-ph]}}.

\bibitem{Capozzi:2021fjo}
F.~Capozzi, E.~Di~Valentino, E.~Lisi, A.~Marrone, A.~Melchiorri, and
  A.~Palazzo, ``{Unfinished fabric of the three neutrino paradigm},''
  \href{http://dx.doi.org/10.1103/PhysRevD.104.083031}{{\em Phys. Rev. D} {\bf
  104} (2021) no.~8, 083031}, \href{http://arxiv.org/abs/2107.00532}{{\tt
  arXiv:2107.00532 [hep-ph]}}.

\bibitem{DayaBay:2018yms}
{\bf Daya Bay} , D.~Adey {\em et al.}, ``{Measurement of the Electron
  Antineutrino Oscillation with 1958 Days of Operation at Daya Bay},''
  \href{http://dx.doi.org/10.1103/PhysRevLett.121.241805}{{\em Phys. Rev.
  Lett.} {\bf 121} (2018) no.~24, 241805},
  \href{http://arxiv.org/abs/1809.02261}{{\tt arXiv:1809.02261 [hep-ex]}}.

\bibitem{RENO:2018dro}
{\bf RENO} , G.~Bak {\em et al.}, ``{Measurement of Reactor Antineutrino
  Oscillation Amplitude and Frequency at RENO},''
  \href{http://dx.doi.org/10.1103/PhysRevLett.121.201801}{{\em Phys. Rev.
  Lett.} {\bf 121} (2018) no.~20, 201801},
  \href{http://arxiv.org/abs/1806.00248}{{\tt arXiv:1806.00248 [hep-ex]}}.

\bibitem{DoubleChooz:2019qbj}
{\bf Double Chooz} , H.~de~Kerret {\em et al.}, ``{Double Chooz $\theta_{13}$
  measurement via total neutron capture detection},''
  \href{http://dx.doi.org/10.1038/s41567-020-0831-y}{{\em Nature Phys.} {\bf
  16} (2020) no.~5, 558--564}, \href{http://arxiv.org/abs/1901.09445}{{\tt
  arXiv:1901.09445 [hep-ex]}}.

\bibitem{Jimenez:2022dkn}
R.~Jimenez, C.~Pena-Garay, K.~Short, F.~Simpson, and L.~Verde, ``{Neutrino
  Masses and Mass Hierarchy: Evidence for the Normal Hierarchy},''
  \href{http://arxiv.org/abs/2203.14247}{{\tt arXiv:2203.14247 [hep-ph]}}.

\bibitem{Denton:2020uda}
P.~B. Denton, J.~Gehrlein, and R.~Pestes, ``{$CP$ -Violating Neutrino
  Nonstandard Interactions in Long-Baseline-Accelerator Data},''
  \href{http://dx.doi.org/10.1103/PhysRevLett.126.051801}{{\em Phys. Rev.
  Lett.} {\bf 126} (2021) no.~5, 051801},
  \href{http://arxiv.org/abs/2008.01110}{{\tt arXiv:2008.01110 [hep-ph]}}.

\bibitem{Miranda:2019ynh}
L.~S. Miranda, P.~Pasquini, U.~Rahaman, and S.~Razzaque, ``{Searching for
  non-unitary neutrino oscillations in the present T2K and NO$\nu $A data},''
  \href{http://dx.doi.org/10.1140/epjc/s10052-021-09227-0}{{\em Eur. Phys. J.
  C} {\bf 81} (2021) no.~5, 444}, \href{http://arxiv.org/abs/1911.09398}{{\tt
  arXiv:1911.09398 [hep-ph]}}.

\bibitem{Chatterjee:2020yak}
S.~S. Chatterjee and A.~Palazzo, ``{Interpretation of NO$\nu$A and T2K data in
  the presence of a light sterile neutrino},''
  \href{http://arxiv.org/abs/2005.10338}{{\tt arXiv:2005.10338 [hep-ph]}}.

\bibitem{Chatterjee:2020kkm}
S.~S. Chatterjee and A.~Palazzo, ``{Nonstandard Neutrino Interactions as a
  Solution to the $NO\nu A$ and T2K Discrepancy},''
  \href{http://dx.doi.org/10.1103/PhysRevLett.126.051802}{{\em Phys. Rev.
  Lett.} {\bf 126} (2021) no.~5, 051802},
  \href{http://arxiv.org/abs/2008.04161}{{\tt arXiv:2008.04161 [hep-ph]}}.

\bibitem{Forero:2021azc}
D.~V. Forero, C.~Giunti, C.~A. Ternes, and M.~Tortola, ``{Nonunitary neutrino
  mixing in short and long-baseline experiments},''
  \href{http://dx.doi.org/10.1103/PhysRevD.104.075030}{{\em Phys. Rev. D} {\bf
  104} (2021) no.~7, 075030}, \href{http://arxiv.org/abs/2103.01998}{{\tt
  arXiv:2103.01998 [hep-ph]}}.

\bibitem{Rahaman:2021leu}
U.~Rahaman, ``{Looking for Lorentz invariance violation (LIV) in the latest
  long baseline accelerator neutrino oscillation data},''
  \href{http://dx.doi.org/10.1140/epjc/s10052-021-09598-4}{{\em Eur. Phys. J.
  C} {\bf 81} (2021) no.~9, 792}, \href{http://arxiv.org/abs/2103.04576}{{\tt
  arXiv:2103.04576 [hep-ph]}}.

\bibitem{Rahaman:2022rfp}
U.~Rahaman, S.~Razzaque, and S.~U. Sankar, ``{A review of the tension between
  the T2K and NO$\nu$A appearance data and hints to new physics},''
  \href{http://arxiv.org/abs/2201.03250}{{\tt arXiv:2201.03250 [hep-ph]}}.

\bibitem{Ellis:2020ehi}
S.~A.~R. Ellis, K.~J. Kelly, and S.~W. Li, ``{Leptonic Unitarity Triangles},''
  \href{http://dx.doi.org/10.1103/PhysRevD.102.115027}{{\em Phys. Rev. D} {\bf
  102} (2020) no.~11, 115027}, \href{http://arxiv.org/abs/2004.13719}{{\tt
  arXiv:2004.13719 [hep-ph]}}.

\bibitem{Ellis:2020hus}
S.~A.~R. Ellis, K.~J. Kelly, and S.~W. Li, ``{Current and Future Neutrino
  Oscillation Constraints on Leptonic Unitarity},''
  \href{http://dx.doi.org/10.1007/JHEP12(2020)068}{{\em JHEP} {\bf 12} (2020)
  068}, \href{http://arxiv.org/abs/2008.01088}{{\tt arXiv:2008.01088
  [hep-ph]}}.

\bibitem{Super-Kamiokande:2016yck}
{\bf Super-Kamiokande} , K.~Abe {\em et al.}, ``{Solar Neutrino Measurements in
  Super-Kamiokande-IV},''
  \href{http://dx.doi.org/10.1103/PhysRevD.94.052010}{{\em Phys. Rev. D} {\bf
  94} (2016) no.~5, 052010}, \href{http://arxiv.org/abs/1606.07538}{{\tt
  arXiv:1606.07538 [hep-ex]}}.

\bibitem{SKNu2020}
Y.~Nakajima, ``{R}ecent results and future prospects from
  {S}uper-{K}amiokande,'' June, 2020.
\newblock \url{https://doi.org/10.5281/zenodo.3959640}.

\bibitem{KamLAND:2013rgu}
{\bf KamLAND} , A.~Gando {\em et al.}, ``{Reactor On-Off Antineutrino
  Measurement with KamLAND},''
  \href{http://dx.doi.org/10.1103/PhysRevD.88.033001}{{\em Phys. Rev. D} {\bf
  88} (2013) no.~3, 033001}, \href{http://arxiv.org/abs/1303.4667}{{\tt
  arXiv:1303.4667 [hep-ex]}}.

\bibitem{T2K:2019efw}
{\bf T2K} , K.~Abe {\em et al.}, ``{Search for light sterile neutrinos with the
  T2K far detector Super-Kamiokande at a baseline of 295 km},''
  \href{http://dx.doi.org/10.1103/PhysRevD.99.071103}{{\em Phys. Rev. D} {\bf
  99} (2019) no.~7, 071103}, \href{http://arxiv.org/abs/1902.06529}{{\tt
  arXiv:1902.06529 [hep-ex]}}.

\bibitem{Wilks:1938dza}
S.~S. Wilks, ``{The Large-Sample Distribution of the Likelihood Ratio for
  Testing Composite Hypotheses},''
  \href{http://dx.doi.org/10.1214/aoms/1177732360}{{\em Annals Math. Statist.}
  {\bf 9} (1938) no.~1, 60--62}.

\bibitem{MINOS:2017cae}
{\bf MINOS+} , P.~Adamson {\em et al.}, ``{Search for sterile neutrinos in
  MINOS and MINOS+ using a two-detector fit},''
  \href{http://dx.doi.org/10.1103/PhysRevLett.122.091803}{{\em Phys. Rev.
  Lett.} {\bf 122} (2019) no.~9, 091803},
  \href{http://arxiv.org/abs/1710.06488}{{\tt arXiv:1710.06488 [hep-ex]}}.

\bibitem{DayaBay:2016qvc}
{\bf Daya Bay} , F.~P. An {\em et al.}, ``{Improved Search for a Light Sterile
  Neutrino with the Full Configuration of the Daya Bay Experiment},''
  \href{http://dx.doi.org/10.1103/PhysRevLett.117.151802}{{\em Phys. Rev.
  Lett.} {\bf 117} (2016) no.~15, 151802},
  \href{http://arxiv.org/abs/1607.01174}{{\tt arXiv:1607.01174 [hep-ex]}}.

\bibitem{Declais:1994su}
Y.~Declais {\em et al.}, ``{Search for neutrino oscillations at 15-meters,
  40-meters, and 95-meters from a nuclear power reactor at Bugey},''
  \href{http://dx.doi.org/10.1016/0550-3213(94)00513-E}{{\em Nucl. Phys. B}
  {\bf 434} (1995)  503--534}.

\bibitem{MINOS:2020iqj}
{\bf MINOS+, Daya Bay} , P.~Adamson {\em et al.}, ``{Improved Constraints on
  Sterile Neutrino Mixing from Disappearance Searches in the MINOS, MINOS+,
  Daya Bay, and Bugey-3 Experiments},''
  \href{http://dx.doi.org/10.1103/PhysRevLett.125.071801}{{\em Phys. Rev.
  Lett.} {\bf 125} (2020) no.~7, 071801},
  \href{http://arxiv.org/abs/2002.00301}{{\tt arXiv:2002.00301 [hep-ex]}}.

\bibitem{Dentler:2018sju}
M.~Dentler, A.~Hern\'andez-Cabezudo, J.~Kopp, P.~A.~N. Machado, M.~Maltoni,
  I.~Martinez-Soler, and T.~Schwetz, ``{Updated Global Analysis of Neutrino
  Oscillations in the Presence of eV-Scale Sterile Neutrinos},''
  \href{http://dx.doi.org/10.1007/JHEP08(2018)010}{{\em JHEP} {\bf 08} (2018)
  010}, \href{http://arxiv.org/abs/1803.10661}{{\tt arXiv:1803.10661
  [hep-ph]}}.

\bibitem{Super-Kamiokande:2014ndf}
{\bf Super-Kamiokande} , K.~Abe {\em et al.}, ``{Limits on sterile neutrino
  mixing using atmospheric neutrinos in Super-Kamiokande},''
  \href{http://dx.doi.org/10.1103/PhysRevD.91.052019}{{\em Phys. Rev. D} {\bf
  91} (2015)  052019}, \href{http://arxiv.org/abs/1410.2008}{{\tt
  arXiv:1410.2008 [hep-ex]}}.

\bibitem{IceCube:2017ivd}
{\bf IceCube} , M.~G. Aartsen {\em et al.}, ``{Search for sterile neutrino
  mixing using three years of IceCube DeepCore data},''
  \href{http://dx.doi.org/10.1103/PhysRevD.95.112002}{{\em Phys. Rev. D} {\bf
  95} (2017) no.~11, 112002}, \href{http://arxiv.org/abs/1702.05160}{{\tt
  arXiv:1702.05160 [hep-ex]}}.

\bibitem{DUNE:2020ypp}
{\bf DUNE} , B.~Abi {\em et al.}, ``{Deep Underground Neutrino Experiment
  (DUNE), Far Detector Technical Design Report, Volume II: DUNE Physics},''
  \href{http://arxiv.org/abs/2002.03005}{{\tt arXiv:2002.03005 [hep-ex]}}.

\bibitem{Hyper-Kamiokande:2018ofw}
{\bf Hyper-Kamiokande} , K.~Abe {\em et al.}, ``{Hyper-Kamiokande Design
  Report},'' \href{http://arxiv.org/abs/1805.04163}{{\tt arXiv:1805.04163
  [physics.ins-det]}}.

\bibitem{Berryman:2015nua}
J.~M. Berryman, A.~de~Gouv\^ea, K.~J. Kelly, and A.~Kobach, ``{Sterile neutrino
  at the Deep Underground Neutrino Experiment},''
  \href{http://dx.doi.org/10.1103/PhysRevD.92.073012}{{\em Phys. Rev. D} {\bf
  92} (2015) no.~7, 073012}, \href{http://arxiv.org/abs/1507.03986}{{\tt
  arXiv:1507.03986 [hep-ph]}}.

\bibitem{Kelly:2017kch}
K.~J. Kelly, ``{Searches for new physics at the Hyper-Kamiokande experiment},''
  \href{http://dx.doi.org/10.1103/PhysRevD.95.115009}{{\em Phys. Rev. D} {\bf
  95} (2017) no.~11, 115009}, \href{http://arxiv.org/abs/1703.00448}{{\tt
  arXiv:1703.00448 [hep-ph]}}.

\end{thebibliography}\endgroup

\end{document}